\documentclass[10pt]{emulateapj}

\usepackage{apjfonts}
\usepackage{graphicx}
\usepackage{natbib}
\usepackage{amsmath}

\begin{document}

\title{The Lifetime and Powers of FR~IIs in Galaxy Clusters}
\shorttitle{FR~IIs in Galaxy Clusters}

\author{
{Joe Antognini}\altaffilmark{1},
{Jonathan Bird}\altaffilmark{1, 2}, and
{Paul Martini}\altaffilmark{1, 2}
}

\shortauthors{Antognini, Bird, \& Martini}

\altaffiltext{1}{Department of Astronomy, The Ohio State University, 140 W
18$^{\rm{th}}$ Ave., Columbus, Ohio 43210, USA}

\altaffiltext{2}{Center for Cosmology \& Astroparticle Physics, The Ohio
State University, 191 W Woodruff Ave., Columbus, Ohio 43210, USA}

\email{antognini@astronomy.ohio-state.edu}

\date{\today}

\begin{abstract}

We have identified and studied a sample of 151 FR~IIs found in brightest
cluster galaxies (BCGs) in the MaxBCG cluster catalog with data from FIRST
and NVSS.  We have compared the radio luminosities and projected lengths of
these FR~IIs to the projected length distribution of a range of mock
catalogs generated by an FR~II model and estimate the FR~II lifetime to be
$1.9 \times 10^8$ yr.  The uncertainty in the lifetime calculation is a
factor of two, due primarily to uncertainties in the ICM density and the FR
II axial ratio.  We furthermore measure the jet power distribution of FR
IIs in BCGs and find that it is well described by a log-normal distribution
with a median power of $1.1 \times 10^{37}$ W and a coefficient of
variation of 2.2.  These jet powers are nearly linearly related to the
observed luminosities, and this relation is steeper than many other
estimates, although it is dependent on the jet model.  We investigate
correlations between FR~II and cluster properties and find that galaxy
luminosity is correlated with jet power.  This implies that jet power is
also correlated with black hole mass, as the stellar luminosity of a BCG
should be a good proxy for its spheroid mass and therefore the black hole
mass.  Jet power, however, is not correlated with cluster richness, nor is
FR~II lifetime strongly correlated with any cluster properties.  We
calculate the enthalpy of the lobes to examine the impact of the FR~IIs on
the ICM and find that heating due to adiabatic expansion is too small to
offset radiative cooling by a factor of at least six.  In contrast, the jet
power is approximately an order of magnitude larger than required to
counteract cooling.  We conclude that if feedback from FR~IIs offsets
cooling of the ICM, then heating must be primarily due to another mechanism
associated with FR~II expansion.

\end{abstract}


\section{Introduction}
\label{sec:intro}

It has long been known that fewer extremely high-mass galaxies are observed
than simple, theoretical calculations predict \citep{white91}.  As random
overdensities from the big bang collapse, hot gas in the largest
overdensities cools and condenses toward their centers to form the
progenitors of brightest cluster galaxies (BCGs).  In the absence of any
source of heating, gas from the protocluster will continue to radiatively
cool onto the BCG at rates as high as $\sim$1000 M$_{\odot}$ yr$^{-1}$ in
so-called ``cooling flows'' (see \citealp{fabian94} for a review).
However, X-ray observations indicate that the gas infall rate at the
centers of most clusters is about 10\% of that predicted by cooling flow
models \citep{peterson01, tamura01}.

It is clear that some mechanism heats gas on the cluster scale roughly
isotropically (see \citealt{mcnamara07} for a recent review).  The best
candidate is some form of radio-mode feedback.  Since BCGs tend to be
disproportionately radio-loud, it has been suggested that most BCGs go
through a radio-loud phase \citep{burns90, best05b}.  In this phase an AGN
emits two radio jets which inflate large ($\sim$100 -- 1000 kpc),
overpressured lobes in the intracluster medium (ICM).  These radio lobes
are often spatially coincident with X-ray cavities, implying that they
physically displace and inject substantial energy into some regions of the
ICM \citep{fabian00, mcnamara00}.  The power required to inflate these
cavities appears to be sufficient to counteract cooling of the hot gas
\citep{birzan04, rafferty06}.  Nevertheless, it remains unclear how this
energy heats the ICM sufficiently isotropically.  Shocks and sound waves
associated with the expansion of the lobe can transfer some heat from the
lobe to the general ICM \citep{jones02, forman05, fabian05} as could cosmic
ray diffusion from the lobes \citep{mathews09}.  Conduction can ameliorate
this problem to some extent, but to properly estimate the heating due to
conduction requires detailed magnetohydrodynamic simulations
\citep{dolag04}. 

Large, extended radio sources have been well-studied for many decades.  An
early morphological classification scheme of double-lobed radio sources was
developed by \citet{fanaroff74}.  Core-brightened sources are classified as
Type I sources (FR~Is) and edge-brightened sources are classified as Type
II sources (FR~IIs).  Although the classification is purely morphological,
FR~IIs are generally more luminous than FR~Is.  The jet power supplied by
the AGN is believed to play an important role in the FR~I/II dichotomy, but
the occasional cases of ``hybrid'' sources which exhibit one FR~I lobe and
one FR~II lobe indicate that environmental factors are not negligible
\citep[e.g.,][]{gopal-krishna00}.  

Because FR~IIs have sharp edges, many important properties (e.g., projected
length, axial ratio, and luminosity) are well defined and independent of
the sensitivity of the observation.  By contrast, FR~Is exhibit long plumes
of fading radio emission, so properties like the length and total
luminosity are more dependent on the sensitivity of the observation.  FR
IIs are therefore easier to study as probes of AGN activity and their
effects on the ICM with radio observations alone.  Furthermore, the advance
speeds and axial ratios of FR~IIs are similar across several orders of
magnitude in lobe length, indicating that FR~II evolution is approximately
self-similar.  This inference has additionally been supported by more
detailed hydrodynamical simulations \citep{carvalho02a, carvalho02b}.  The
assumption of self-similarity has allowed the possibility of analytical or
semi-analytical models to describe FR~II evolution.  A number of such
models have been developed over the past two decades.  Among the most
complete and most widely used is the model presented in \citet{kaiser97a}
and \citet*{kaiser97b}, henceforth referred to as the KDA model.  We use
the KDA model exclusively throughout this paper. 

The calculation of the length evolution of the lobe is straightforward and
is similar in all FR~II models.  Earlier FR~II studies have used FR~II lobe
length distributions to estimate FR~II ages of $10^7$ to $10^8$ years
\citep{blundell99, bird08}.  These measurements are in reasonable agreement
with spectral aging estimates and measurements of the buoyant rising times
of ghost cavities, which find FR~II ages in the range of 10$^6$ to 10$^8$
yr \citep{allen06, o'dea09}.  These estimates generally only reflect the
current age of individual sources, or subsets of the population that may
not be representative, and do not provide a direct estimate of the lifetime
or energy of typical sources.  While it is possible to infer an FR~II
lifetime from an age distribution, it requires good knowledge of the
selection efficiency as a function of age.  Studies which attempt to infer
the FR~II lifetime from length distributions can constrain their selection
biases more easily \citep{blundell99, bird08}, but still suffer from
uncertainties in the jet power.  

The FR~II jet power is a difficult quantity to determine and the few FR~IIs
for which it has been estimated have tended to be unusually bright, and
hence, high-power sources \citep{machalski04a, mcnamara07}.  Inferring the
general FR~II jet power distribution from the high-power end alone,
however, requires substantial and uncertain extrapolation.  Nevertheless,
recent studies have begun to probe the low-power end of the FR~II jet power
distribution and have found typical jet powers of $10^{37}$ W and jet
powers as low as $10^{36}$ W \citep{cavagnolo10, ito08, punsly11}.  These
estimates place the median FR~II power lower than older estimates by at
least an order of magnitude.  Although FR~II lifetime estimates do not
scale sensitively with FR~II power, a substantial overestimate of the jet
power will lead to an underestimate of the age or lifetime.

In this work we measure the ages and powers of 151 FR~IIs in BCGs from the
MaxBCG galaxy cluster catalog of \citet{koester07b}.  We simultaneously fit
the length and luminosity of the FR~IIs with the KDA model and derive an
estimate of the power distribution.  We then obtain an improved lifetime
and duty cycle estimate and examine the importance of FR~IIs in ICM
heating.  The paper is organized as follows: in \S\ref{sec:sample} we
describe the sample selection algorithm, the luminosity and projected
length measurements, and list general properties of the FR~IIs in our
sample; in \S\ref{sec:model} we motivate our use of the KDA model, provide
a summary of the model's important features, and describe our technique to
estimate the jet powers; in \S\ref{sec:frii_lifetime} we fit the observed
length distribution to a set of mock catalogs generated by a Monte Carlo
simulation of FR~II observations; in \S\ref{sec:correlations} we
investigate correlations between FR~II power and lifetime with various
cluster properties and examine the FR~II fraction as a function of cluster
properties; in \S\ref{sec:discussion} we compare our results with those in
the literature and discuss the heating effects of FR~IIs on the ICM; we
summarize our results in \S\ref{sec:summary}.

Throughout the paper we assume a $\Lambda$CDM cosmology with $\Omega_m$ =
0.27, $\Omega_{\Lambda}$ = 0.73, and $H_0$ = 70 km s$^{-1}$ Mpc$^{-1}$.

\section{Sample Selection}
\label{sec:sample}

Our sample selection process is similar to the process used in
\citet{bird08}, namely we identify FR~IIs hosted by central galaxies in
dense environments.  The major difference is that we use the MaxBCG catalog
of \citet{koester07b} to identify BCGs rather than the catalog of galaxy
groups published by \citet{berlind06}.  We describe properties of the
galaxy cluster catalog in \S\ref{subsec:cluster_catalog} and properties of
the catalogs of radio sources in \S\ref{subsec:radio_catalog}.  Our process
for cross-correlating the two catalogs is described in
\S\ref{subsec:crosscorr}.  \S\ref{subsec:sample_measurements} describes the
sample selection method and properties of the final FR~II sample.

\subsection{Cluster Catalog}
\label{subsec:cluster_catalog}

The MaxBCG catalog consists of 13,823 galaxy clusters selected from the
Sloan Digital Sky Survey with the maxBCG red-sequence method.  The red
sequence method attempts to find BCGs by selecting galaxies lying on the
E/S0 ridgeline (a region of color-magnitude space where old, passively
evolving galaxies are found) which are also brighter than their neighboring
galaxies.  The algorithm identifies potential BCGs in an overdensity of
galaxies with similar colors and redshifts.  The red sequence method is
described in detail in \citet{koester07b}.  Tests using mock catalogs
performed by \citet{koester07a} indicate that the MaxBCG catalog is at
least 90\% pure and 85\% complete for clusters larger than $10^{14}$
M$_{\odot}$.

The number of galaxies with similar colors and redshifts in the overdense
region, $N_{\textrm{gal}}$, is an estimate of the total number of
galaxies in the cluster.  $N_{\textrm{gal}}$ is later refined to
$N_{\textrm{gal}}^{R200}$, the number of galaxies within the
radius at which the galaxy density is 200 times larger than the mean
density of galaxies with $-24 \leq M_r \leq -16$ mag.  Throughout this paper we
use $N_{\textrm{gal}}^{R200}$ as a proxy for the cluster richness.

\begin{figure*}
\centering
\includegraphics[width=16cm]{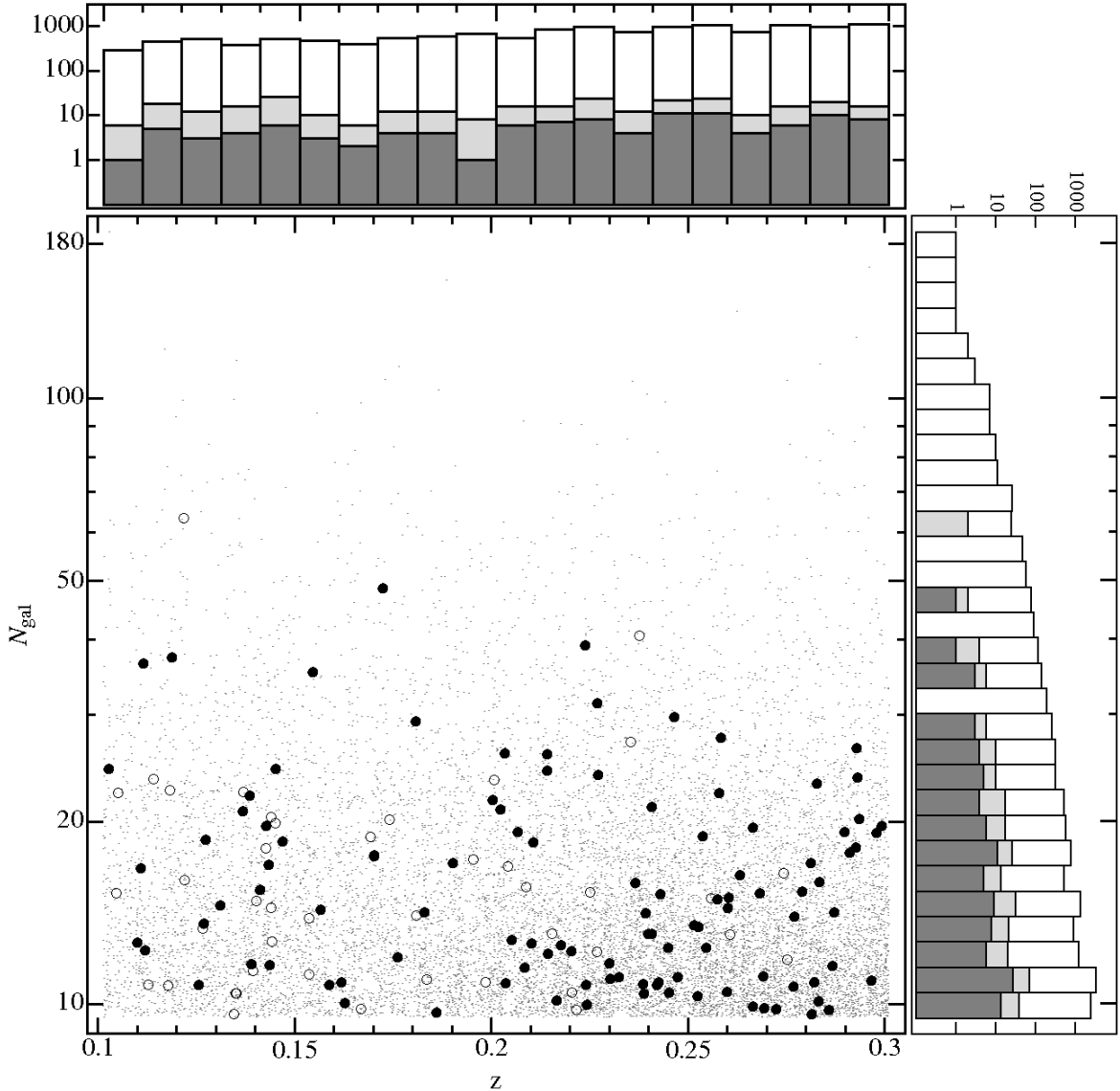}

\caption{The distribution of cluster richness and redshift in the MaxBCG
catalog (small gray points), the full FR~II sample (large open and filled
circles), and the volume-limited FR~II sample (filled circles alone).
Histograms of the cluster richness and redshift are also shown.  The
unfilled area represents the counts for all clusters in the MaxBCG catalog,
the light gray area the counts for those clusters with FR~IIs in our full
sample, and the dark gray area the counts for those clusters with FR~IIs in
our volume-limited sample.}\label{fig:z_ngal}

\end{figure*}

The MaxBCG cluster catalog spans a range in photometric redshift of $z =
0.1 - 0.3$ and a range in richness of $N_{\textrm{gal}}^{R200} =
10 - 188$.  39\% of the BCGs in the MaxBCG catalog also have spectroscopic
redshifts and the dispersion between the spectroscopic and photometric
redshifts is 0.01.  Throughout this paper we use spectroscopic redshifts
whenever they exist and otherwise use photometric redshifts.  The
distribution of cluster richness and redshifts in the MaxBCG catalog is
shown in Figure \ref{fig:z_ngal}.

Although the MaxBCG catalog does not provide cluster mass estimates, the
clusters span a range of $r$-band luminosity from $7 \times 10^{10}$
L$_{\odot}$ to $3 \times 10^{12}$ L$_{\odot}$ with a median
luminosity of $2 \times 10^{11}$ L$_{\odot}$.  For a typical cluster
mass-to-light ratio of $\sim$350 M$_{\odot} /$ L$_{\odot}$
\citep{carlberg97, sheldon09}, the typical cluster mass in the MaxBCG
catalog is $\sim$7 $\times 10^{14}$ M$_{\odot}$ with a spread of about
one order of magnitude in either direction.

\subsection{Radio Catalogs}
\label{subsec:radio_catalog}

We selected radio sources associated with the BCGs of these galaxy clusters
using the July 16, 2008 version of the FIRST radio catalog \citep{white97}.
The FIRST radio catalog consists of 816,331 radio sources identified from
9055 deg$^2$ observed as part of the FIRST radio survey at 1.4 GHz
\citep{becker95}.  Data for this version of the catalog were obtained with
the Karl G.~Jansky Very Large Array (VLA) from 1993 through 2004.  The
angular resolution of the FIRST survey is $5.\!\!^{\prime\prime} 4$ and the
typical rms flux noise is 0.15 mJy.  Dirty beams from the VLA were CLEANed
to produce images for the FIRST survey with $1.\!\!^{\prime\prime} 4$
pixels. 

The FIRST radio catalog was produced from the survey data using a
specialized source extraction algorithm dubbed HAPPY.  HAPPY searches for
local maxima and attempts to fit them to up to four elliptical Gaussian
components.  Fits which produce physical results and are not too close to
an image edge are accepted into the catalog.  The catalog is then refined
to eliminate duplicate and spurious sources.  To be considered a detection,
a source must be at least five times larger than the rms noise.  The
minimum flux of any source is therefore typically above 0.75 mJy. 

The resulting catalog provides, among other parameters, the coordinates,
the local rms noise, the peak and integrated flux, and a warning flag if
the source is a potential sidelobe of a nearby bright source.  The two flux
measurements are corrected for CLEAN bias \citep{becker95, condon98}.

Because the FIRST survey has relatively high resolution and treats extended
sources as collections of point-like sources, much of the flux of extended
objects can be resolved out and missed, thereby leading to underestimates
of the total flux.  Since FR~IIs are extended, we complement our flux
measurements of sources from the FIRST catalog using the NRAO VLA Sky
Survey (NVSS).  NVSS is a similar radio survey to FIRST, but has an angular
resolution of $45^{\prime \prime}$ rather than $5.\!\!^{\prime \prime}4$,
thereby providing more accurate flux measurements for extended sources
\citep{condon98}.  The flux limit for the NVSS catalog is $S \simeq 2.5$
mJy.  Although this is larger than FIRST by a factor of approximately
three, extended sources are broken into fewer components in the NVSS
catalog and we lose no FR~IIs in the NVSS catalog.

\subsection{Cross-Correlating the Two Catalogs}
\label{subsec:crosscorr}

We first selected a sample of FR~II candidates by identifying radio sources
between $10.\!\!^{\prime\prime}8$ and $3.4^{\prime}$ of every BCG.  The
upper bound of $3.4^{\prime}$ was chosen to be small enough to limit the
number of spurious candidates, yet large enough that real FR~IIs would not
be eliminated.  At a redshift of $z = 0.1$, the lowest in our sample, this
cut would eliminate any FR~IIs longer than 370 kpc.  Since we found no FR
IIs longer than 287 kpc, we conclude that this choice of maximum angular
separation has no effect on our sample size.  The lower bound of
$10.\!\!^{\prime\prime}8$ was chosen so that the FR~IIs would be at least a
factor of two greater than the FIRST resolution.  The selection function
for short FR~IIs near the FIRST resolution limit is difficult to
characterize; some FR~IIs near the FIRST resolution limit were discovered
by our selection algorithm, whereas others, due to the chance peculiarities
of their morphology or surroundings, may have been missed.  By selecting
only FR~IIs whose lobes each subtend at least two resolution elements of
FIRST, we ensure that our selection algorithm only identifies FR~IIs that
are resolved by FIRST. 

Except in rare cases when a radio jet is oriented very nearly along the
line of sight and has one lobe strongly beamed towards the Earth and the
other lobe strongly beamed away, radio jets appear to be extremely
symmetrical \citep{scheuer95}.  We thus identified FR~II candidates as
radio sources that were equidistant from the central BCG to within 30\% and
at least $135^{\circ}$ away from each other relative to the BCG.  These
selection criteria yielded an initial sample of 617 FR~II candidates.  We
then superimposed FIRST contours on SDSS images of these FR~II candidates
and inspected them by eye to eliminate contaminants.  There were two major
sources of contamination: (1) pairs of radio galaxies on opposite sides of
the BCG or (2) a radio galaxy paired with a faint radio source unassociated
with any optical counterpart.  Aside from their spatial coincidence with
optical sources, contaminants in both cases could be easily distinguished
from faint FR~II lobes due to the relatively large asymmetry between the
position of the two sources with respect to the BCG, the relatively large
asymmetry in flux, and the lack of any extended emission toward the BCG.
We also removed 31 obvious FR~Is from the sample.  This resulted in a
preliminary sample of 432 FR~IIs.  Less-obvious FR~I sources were removed
in a more quantitative second pass described in the following section.

\subsection{FR~II Classification and Sample Selection}
\label{subsec:sample_measurements}

Although FR~Is can often be distinguished from FR~IIs based on luminosity
alone, the formal division between FR~I and FR~II sources is purely
morphological \citep{fanaroff74}.  Sources for which the hotspot is closer
to the tip of the lobe are classified as FR~IIs, and sources for which the
hotspot is closer to the central galaxy are classified as FR~Is.  Except in
extreme cases, since projection effects cannot drastically change the ratio
between the distance to the hotspot and the distance to the lobe tip, the
projected positions of the hotspot, lobe tip, and central galaxy are
sufficient to eliminate any remaining FR~Is from our sample.

We therefore measured the position of the hotspot and the tip of the lobe
for each FR~II in the sample.  We identified the position of the hotspot as
the centroid at the local maximum near the FR~II's edge.  It is inherently
more difficult to determine the tip of the lobe.  Although the lobes of FR
II sources have better-defined edges relative to FR~Is, the tip of a lobe
is nevertheless not entirely well-defined due to variations in the
sensitivity of the radio data and the often complex morphology of the
source.  We defined the tip of the lobe as the most distant pixel from the
central BCG that was contiguous with the main lobe and above the FIRST
detection threshold by five times the rms noise.  Once we determined the
position of the hotspot and tip of the lobe, we removed less-evident FR~Is
from the sample.  

We derive the projected physical length of each source from the separation
of each lobe tip from the BCG and the angular diameter distance to the
cluster.  The errors on the projected source sizes are dominated by random
errors in the photometric redshift and the uncertainty in the definition of
the lobe tip.  We estimate that the total uncertainty in the physical lobe
length is $\lesssim5\%$.  As shall be shown in
\S\S\ref{subsec:axial_ratio_dependence} and~\ref{subsec:icm_dependence},
the uncertainty in our lifetime calculation is dominated by uncertainty in
the ICM and the FR~II model, rather than the uncertainty in physical lobe
length.

\begin{figure}
\centering
\includegraphics[width=8cm]{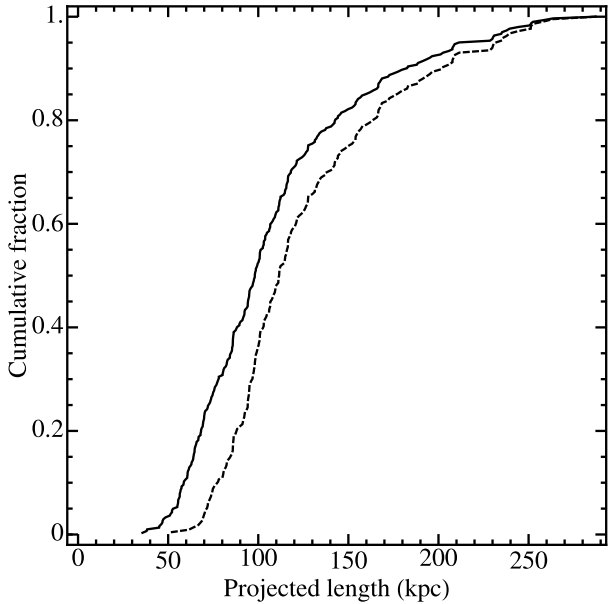}

\caption{The length distribution of the full (solid line) and the
volume-limited (dashed line) FR~II samples.  Since shorter sources are not
resolved by FIRST at high redshifts, the volume-limited sample contains
fewer short FR~IIs than the flux-limited sample.\label{fig:raw_cut_length}}

\end{figure}

We obtained the fluxes for the FR~IIs from the NVSS catalog.  Because NVSS,
like FIRST, breaks up extended sources into discrete components, the flux
from each component was summed to give the total flux from each side of the
FR~II source.  Since NVSS cannot resolve the double-lobed morphology of the
smaller sources in the sample, it was often impossible to assign a separate
flux to each lobe.  In these instances we measured the total flux from both
lobes.  When fluxes from individual lobes were needed, such as when
estimating the power and age of the jet as described in
\S\ref{subsec:q_measurement}, we simply assigned each lobe half the total
flux.  Since this prescription requires that both lobes be FR~IIs, we made
a final cut in our sample to eliminate any radio sources that consisted of
an FR~II lobe paired with an FR~I lobe.  This cut eliminated 42 nominal FR
I/II mixed sources.  We note that the vast majority of these mixed sources
are not ``true'' FR~I/II hybrid sources in the usual sense of a
well-defined FR~I plume on one side and a well-defined FR~II lobe on the
other (so-called ``HYMOR'' sources, \citealt{gopal-krishna00}).  Rather, in
these sources the hotspots of both lobes are approximately halfway along
the length of the lobe and one hotspot is slightly beyond halfway to the
lobe tip and the other slightly before.  Application of all of these
selection criteria resulted in a flux-limited sample of 151 FR~II lobe
pairs.  We perform the bulk of analysis with this sample and will simply
refer to it as ``the sample'' or the ``full sample'' rather than the
flux-limited sample.  The distribution of the FR~IIs in our sample in
redshift and host cluster richness is shown in Figure \ref{fig:z_ngal} and
the projected length distribution is shown in Figure
\ref{fig:raw_cut_length}.  

This sample of 151 FR~II lobe pairs suffers from several significant
selection biases.  Specifically, intrinsically longer jets are easier to
resolve at larger distances, and more luminous FR~IIs are easier to detect
at larger distances.  Since more powerful FR~IIs are both longer and more
luminous at a given age than less powerful FR~IIs \citep{kaiser97b}, this
sample is biased towards more powerful FR~IIs at higher redshifts.  These
biases become problematic when attempting to determine whether FR~II
lifetime or duty cycle change with redshift.  To ameliorate this problem,
we introduced a cut into our sample that required every FR~II source meet
our selection criteria at $z = 0.3$, the largest redshift of the clusters
in our sample.  That is, if the source were at $z = 0.3$, the angular
separation between the two lobes of the jet would have to be at least
$10.\!\!^{\prime \prime}$8, and the flux from the FIRST component for both
hotspots would have to be above the FIRST detection limit of five times the
rms background noise.  There are 108 pairs of FR~II lobes in this sample,
which we will refer to as the volume-limited sample.  The length
distribution of the volume-limited sample is also shown in
Figure~\ref{fig:raw_cut_length}.  

There is likely a slight overcorrection in the volume-limited sample, but
one that is difficult to quantify.  Because extended sources appear more
compact at larger distances, high-resolution surveys like FIRST can collect
more flux from the source.  Thus if an extended source at $z < 0.3$ is
moved out to $z = 0.3$, the flux detected by FIRST would be slightly larger
than one would expect based on the change in distance alone.  Since this
effect is difficult to characterize accurately due to the complex
morphology of the sources, we simply note that low-redshift sources in our
volume-limited sample are slightly biased towards larger lengths.

\begin{figure*}
\centering
\includegraphics[width=16cm]{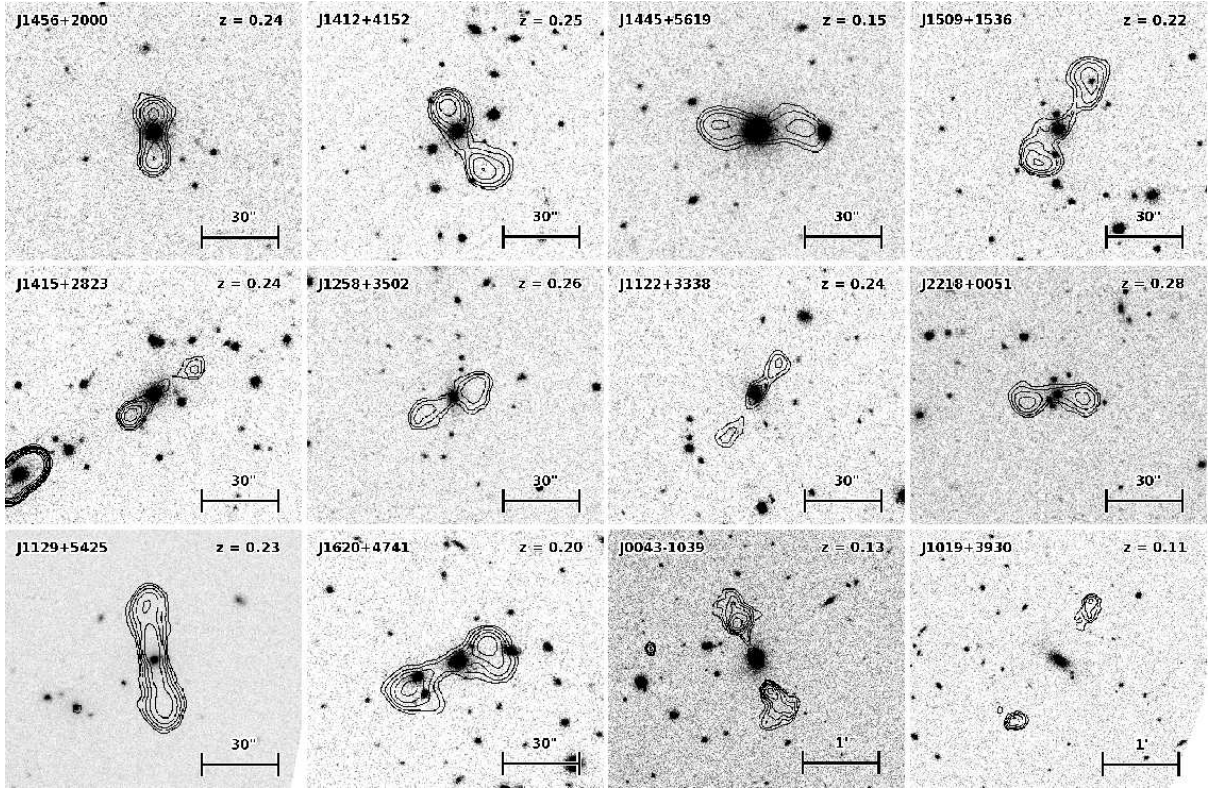}

\caption{Twelve randomly chosen FR~II radio sources from the sample.  These
are SDSS $r$-band images with contours from the FIRST survey
superimposed.\label{fig:mosaic}}

\end{figure*}

\begin{figure}
\centering
\includegraphics[width=8cm]{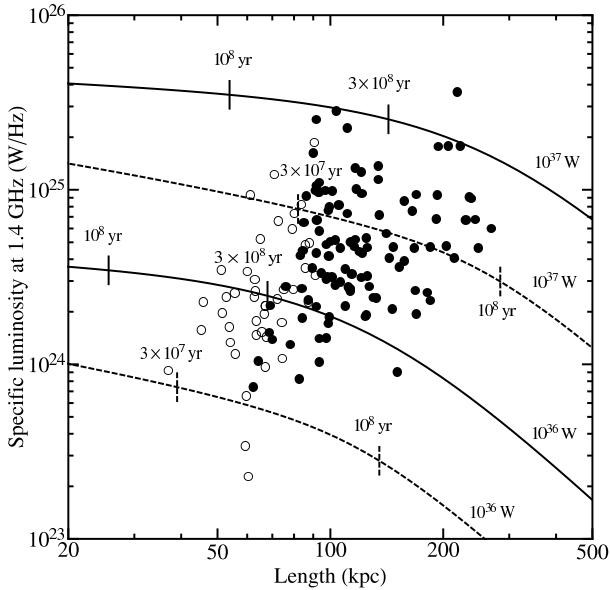}

\caption{Radio luminosity vs.~lobe length for FR~IIs in the full (open and
filled circles) and volume-limited samples (filled circles only).  There is
no clean line in length-luminosity space that divides the two samples
because the sample cut was made with the length to the hotspot, while the
length to the edge of the lobe is shown in this figure and the ratio
between the hotspot and lobe distance can vary by up to a factor of two.
For comparison, evolutionary tracks from the KDA model for four extreme
cases are shown.  The solid lines show an ICM model with $a = 260$ kpc and
$\rho = 5.9 \times 10^{-24}$ kg/m$^3$, characteristic of a rich cluster,
and the dashed lines show an ICM model with $a = 2$ kpc and $\rho = 7.2
\times 10^{-22}$ kg/m$^3$, characteristic of a small group.  Both the upper
solid and dashed lines represent a jet with a power of $Q = 10^{37}$ W, and
the lower solid and dashed lines represent a jet with a power of $Q =
10^{36}$ W.  The left and right ticks of the solid lines represent ages of
$10^8$ and $3 \times 10^8$ yr, respectively, and the left and right ticks
of the dashed lines represent ages of $3 \times 10^7$ and $10^8$ yr,
respectively.\label{fig:pd_fig}}

\end{figure}

The properties of all of the FR~II sources and their host clusters are
listed in Table \ref{tbl:frii_sample}.  Figure~\ref{fig:mosaic} displays
examples of several randomly selected FR~IIs from our sample and Figure
\ref{fig:pd_fig} presents a $P$-$D$ diagram of the full sample.  The
sources span a range of lengths of $48 - 287$ kpc with a median length of
95 kpc.  The host clusters span a range of richness from 10 (the minimum
richness in the MaxBCG catalog) to 69 with a median richness of 14 and the
full redshift range of the MaxBCG catalog of $z = 0.1 - 0.3$.  The sample
spans a range in specific luminosity of $4.4 \times 10^{23} - 7.2 \times
10^{25}$ W m$^{-2}$ Hz$^{-1}$ with a median luminosity of $8.0 \times
10^{24}$ W m$^{-2}$ Hz$^{-1}$ (all at an observed frequency of 1.4 GHz).
Note that we do not $K$-correct the luminosities of the FR~II sample at
this point, but account for $K$-corrections in the mock catalogs (described
in \S\ref{subsec:mock_generation}).  

These luminosities are smaller by at least an order of magnitude than the
luminosities of FR~II sources in many earlier studies
\citep[e.g.,][]{laing83, subrahmanyan96, cotter96} and are smaller than the
luminosities of well-studied FR~IIs like Cygnus A, 3C 47, and 3C 295 by
over three orders of magnitude \citep{braude69}.  Indeed, 75\% of the
sources in our sample fall below the FR~I/II demarcation line set out in
\citet{ledlow96}.  While it is possible that
some sources close to our resolution cutoff may be FR~Is confused for FR
IIs, the vast majority of our sources are well resolved.  Specifically,
$\sim$75\% of our sample has an angular separation of at least three FIRST
beams, $\sim$50\% of our sample has an angular separation of at least four
FIRST beams, and $\sim$25\% of our sample has an angular separation of at
least six FIRST beams.  

The relatively low luminosity of the FR~IIs in our sample is due to the
FIRST survey's superior sensitivity and resolution with respect to earlier
radio surveys.  Probing the faint end of the FR~II luminosity function
allows us to better constrain the lower-limit of FR~II jet power and study
the characteristics of more typical FR~IIs.  We address this subject in
detail in \S\ref{subsec:q_measurement}.

\begin{deluxetable*}{lccccccccccc}
\tablewidth{0pt}
\tablecolumns{12}
\tablecaption{FR~II Sample}
\tablehead{
\colhead{SDSS ID} &
\colhead{$z$} &
\colhead{$N_{200}$} &
\colhead{$L_{r, \, \textrm{BCG}}$} &
\colhead{$L_{z, \, \textrm{BCG}}$} &
\colhead{$F_{\textrm{NVSS}}$} &
\colhead{$P_{\textrm{NVSS}}$} &
\colhead{$l_1$} &
\colhead{$l_2$} &
\colhead{$l_1$} &
\colhead{$l_2$} &
\colhead{In volume-} \\
\colhead{} &
\colhead{} &
\colhead{} &
\colhead{($10^{10} \rm{L}_{\odot}$)} &
\colhead{($10^{10} \rm{L}_{\odot}$)} &
\colhead{(mJy)} &
\colhead{($10^{24}$ W Hz$^{-1}$)} &
\colhead{(kpc)} &
\colhead{(kpc)} &
\colhead{(arcsec)} &
\colhead{(arcsec)} &
\colhead{limited sample?} \\
\colhead{(1)} &
\colhead{(2)} &
\colhead{(3)} &
\colhead{(4)} &
\colhead{(5)} &
\colhead{(6)} &
\colhead{(7)} &
\colhead{(8)} &
\colhead{(9)} &
\colhead{(10)} &
\colhead{(11)} &
\colhead{(12)} \\
}

\startdata

J001247.6+004715.8 & 0.154 & 14 & 5.46 & 6.83 & 61.7 & 3.9 & 46.9 & 56.0 & 13.2 & 15.8 & No \\
J003614.2-090451.8 & 0.270 & 10 & 4.43 & 5.24 & 17.0 & 3.8 & 121.0 & 127.8 & 18.1 & 19.1 & Yes \\
J004312.9-103956.1 & 0.128 & 19 & 7.71 & 9.60 & 148.4 & 6.2 & 119.6 & 122.8 & 41.4 & 42.5 & Yes \\
J013157.8-081955.0 & 0.140 & 11 & 7.51 & 9.65 & 469.0 & 24.2 & 55.5 & 85.9 & 17.3 & 26.8 & No \\
J073826.2+451719.1 & 0.221 & 10 & 6.36 & 8.07 & 91.9 & 13.0 & 65.3 & 80.3 & 12.3 & 15.1 & No \\
J074616.9+220203.5 & 0.260 & 19 & 6.16 & 7.51 & 30.4 & 6.2 & 86.1 & 110.1 & 13.5 & 17.2 & Yes \\
J075457.8+210129.0 & 0.256 & 14 & 5.42 & 6.78 & 21.7 & 4.3 & 64.2 & 74.0 & 10.8 & 11.8 & Yes \\
J075614.8+251340.4 & 0.202 & 11 & 8.67 & 10.77 & 188.8 & 21.7 & 75.5 & 111.1 & 15.7 & 23.2 & Yes \\
J080107.0+175845.3 & 0.146 & 24 & 5.05 & 6.52 & 890.7 & 49.9 & 78.1 & 104.8 & 23.3 & 31.3 & Yes \\
J080641.4+494628.4 & 0.245 & 13 & 5.90 & 7.66 & 84.2 & 15.0 & 84.9 & 112.4 & 14.2 & 18.8 & Yes \\

\enddata

\tablecomments{FR~II sources in in our sample ordered by right ascension.
(1): SDSS identifier.  (2): BCG redshift.  (3): Cluster richness.  (4, 5):
The $r$- and $z$-band luminosities of the BCG, respectively.  (6): Total
flux as measured by NVSS in mJy.  (7): Specific radio luminosity of the FR
II calculated from the NVSS flux in $10^{24}$ W Hz$^{-1}$.  (8, 9):
Projected length of the two lobes in kpc.  (10, 11): Angular separation of
the two lobes in arcseconds.  (12): Identifies whether the FR~II is also in
the volume-limited sample.  Table \ref{tbl:frii_sample} is presented in its
entirety in the electronic edition.  A portion is shown here for guidance
regarding its form and content.}\label{tbl:frii_sample}

\end{deluxetable*}


\section{Modeling the length distribution}
\label{sec:model}

We estimate the lifetime and jet powers of FR~IIs with detailed models of
their luminosity evolution, environments, and the sample selection.  This
approach requires a good model of FR~II length and luminosity evolution,
and our choice is discussed in \S\ref{subsec:model_properties}.  The model
for the ICM can impact the lifetime determination as well, so we discuss
the properties of the ICM models we use in \S\ref{subsec:model_icm}.  In
\S\ref{subsec:q_measurement} we introduce the method by which we estimate
FR~II jet powers, a major contribution to the uncertainty in previous FR~II
lifetime measurements.  We present a relationship between the jet power and
the radio luminosity in \S\ref{subsec:pjet_lum}.  We finally discuss our
use of mock catalogs to account for selection effects in
\S\ref{subsec:mock_generation}. 

\subsection{Properties of the FR~II Model}
\label{subsec:model_properties}

To determine the FR~II source lifetime we need an accurate, analytic model
of the length and luminosity evolution of the sources.  A number of
analytic models of FR~II source evolution exist in the literature
\citep[e.g.,][]{bicknell97, kaiser97b, blundell99, manolakou02, kino05}.
The Bicknell, KDA, Blundell, and Manolakou models all assume self-similar
growth of the jet.  Although this assumption appears to be approximately
correct, hydrodynamical simulations by \citet{carvalho02a, carvalho02b}
show that it fails in detail in low Mach number jets in ICMs with
relatively flat density profiles.  The more recent model by \citet{kino05}
avoids the assumption of self-similarity, although it only describes the
total kinetic power rather than the radio luminosity of the source.  Since
the density profile of the ICM of galaxy clusters tends to be steep and the
jets in FR~IIs have large Mach numbers, the self-similarity assumption is
not problematic for our purposes.  The radio luminosity of the FR~II
sources plays an important role in our determination of their lifetimes, so
we find the Kino model to be unsuitable for our study.  \citet{barai06}
compared the Bicknell, Manolakou, and KDA models to FR~II radio galaxies in
the 3CRR, 6CE, and 7CRS radio surveys and found that the KDA model best
matched data in the $P$-$D$-$z$-$\alpha$ plane.  Based on this result, and
for purposes of comparison with a similar study by \citet{bird08}, we use
the KDA model to describe the length and luminosity evolution of jets in
our sample. 

The KDA model was recently reworked into a simpler form by
\citet{kaiser07}; we adopt their notation for this paper.  We note,
however, that we have made one modification to the original KDA model.  In
a subsequent paper, \citet{kaiser99b} performed hydrodynamical simulations
to test the accuracy of the KDA model.  While the analytic model agreed
with hydrodynamical simulations, Kaiser \& Alexander found that an
approximation made in \citet{kaiser97a} of
\begin{equation}
\label{eq:assumption}
\frac{p_h}{p_c} = A^2
\end{equation}
overestimates $p_h / p_c$, where $p_h$ is the pressure at the head of the
jet, $p_c$ is the pressure within the cocoon, and $A$ is the axial ratio.
Kaiser (2000) provided an empirical parameterization of this ratio based on
earlier simulations, giving
\begin{equation}
\label{eq:empirical_reformulation}
\frac{p_h}{p_c} = (2.14 - 0.52 \beta) \left( \frac{A}{2} \right)^{2.04 -
0.25 \beta},
\end{equation}
where $\beta$ is the power law slope of the ICM density profile.  This
modification has the following effect on the KDA model.  The length of the
lobe in the original KDA model is given by 
\begin{equation}
\label{eq:length}
D = c_1 \left( \frac{Q}{\rho a^{\beta}} \right)^{1/(5 - \beta)} t^{3 / (5 -
\beta)},
\end{equation}
where $c_1$ is
\begin{equation}
\label{eq:proportionality_constant}
c_1 = \left[ \frac{A^4}{18 \pi} \frac{(\Gamma_x + 1)(\Gamma_l - 1)(5 -
\beta)^3}{9 [\Gamma_l + (\Gamma_l - 1) A^2 / 2] - 4 - \beta} \right]^{1 /
(5 - \beta)},
\end{equation}
$t$ is the age of the FR~II, $Q$ is the jet power, and $\Gamma_x$ and
$\Gamma_l$ are the adiabatic indices of the ICM and the lobe, respectively.
By substituting Equation (\ref{eq:assumption}) for
Equation~(\ref{eq:empirical_reformulation}), the constant $c_1$ becomes
\begin{multline}
\label{eq:final_equation}
c_1^{5 - \beta} = \left( \frac{(2.14 - 0.52 \beta) A^2}{18 \pi}
\left(\frac{A}{2}\right)^{2.04 - 0.25 \beta} \right) \times \\
\left(\frac{(\Gamma_x + 1)(\Gamma_l - 1)(5 - \beta)^3}{9 [ \Gamma_l + 
(\Gamma_l - 1) A^2 / 2] - 4 - \beta} \right).
\end{multline}
For our choice of $\beta = 1.9$ (discussed in \S\ref{subsec:model_icm}),
this modification shortens the predicted lengths by approximately 40\% and
increases the age by approximately 70\%. 

The salient feature of the KDA model is that it predicts the source length
and specific luminosity of a jet given an age, jet power, redshift, and a
number of additional parameters that describe the jet's shape and
environment.  The full list of inputs and several of the most relevant
outputs of the KDA model is displayed in Table \ref{tbl:kda_params} along
with the range over which we vary the parameters and our default values in
this paper.  Because the KDA model must be run several million times for a
given choice of inputs in order to accurately model the length
distribution, it is too computationally expensive to sample the entire
range of parameter space that the KDA model offers.  \citet{bird08} found,
however, that most of the parameters have only a negligible impact on FR~II
lifetime calculations over physically plausible ranges.  We therefore limit
our exploration of parameter space to only those parameters which Bird et
al.~(2008) found to have a substantial impact on lifetime calculations,
namely the density parameter $\rho a^{\beta}$ and the axial ratio $A$. 

\begin{deluxetable*}{lccccc}
\tablewidth{0pt}
\tablecolumns{6}
\tablecaption{Input \& Output Parameters of the KDA
Model}
\tablehead{
\colhead{Parameter type} &
\colhead{Parameter} &
\colhead{Default value} &
\colhead{Minimum value} &
\colhead{Maximum value} &
\colhead{Description} \\
}

\startdata
Primary input & $t$ & \ldots & \ldots & \ldots & FR~II age \\
Primary input & $Q$ & \ldots & \ldots & \ldots & Jet power \\
Primary input & $z$ & \ldots & \ldots & \ldots & Redshift \\
Secondary input & $A$ & 4.0 & 2.0 & 16.0 & Axial ratio (see \S\ref{subsec:axial_ratio_dependence} for our definition of $A$) \\
Secondary input & $\rho$ & $7.2 \times 10^{-26}$ kg m$^{-3}$ & $7.2 \times 10^{-26}$ kg m$^{-3}$ & $7.2 \times 10^{-22}$ kg m$^{-3}$ & Density at $a_0$ \\
Secondary input & $a_0$ & 391 kpc & 2 kpc & 391 kpc & Scale radius of ICM density profile \\
Secondary input & $\beta$ & 1.9 & \ldots & \ldots & Power-law slope of ICM density profile \\
Secondary input & $m$ & 2.14 & \ldots & \ldots & Power-law slope of injection energy of particles in jet \\
Secondary input & $\nu$ & 1.4 GHz & \ldots & \ldots & Observation frequency \\
Secondary input & $\Gamma_x$ & 5/3 & \ldots & \ldots & Adiabatic index of the ICM \\
Secondary input & $\Gamma_l$ & 4/3 & \ldots & \ldots & Adiabatic index of the lobe \\
Secondary input & $\Gamma_B$ & 4/3 & \ldots & \ldots & Adiabatic index of the magnetic field energy density \\
Secondary input & $\gamma_{\textrm{min}}$ & 1 & \ldots & \ldots & Minimum Lorentz factor for electrons in the jet \\
Secondary input & $\gamma_{\textrm{max}}$ & $10^{10}$ & \ldots & \ldots & Maximum Lorentz factor for electrons in the jet \\
Secondary input & $k$ & 0 & \ldots & \ldots & Ratio of energy stored in non-radiating particles \\
 & & & & & to the sum of the energy in the magnetic field and the \\
 & & & & & relativistic electrons. \\
Output & $l$ & \ldots & \ldots & \ldots & FR~II Lobe length \\
Output & $L$ & \ldots & \ldots & \ldots & FR~II radio luminosity \\
Output & $p_c$ & \ldots & \ldots & \ldots & FR~II lobe pressure \\
\enddata

\tablecomments{We designate input parameters which are likely to be
significantly different among FR~IIs to be ``Primary inputs'' and therefore
assign no default values to them.  A similar study by \citet{bird08} found
that variation of most of the secondary inputs does not substantially
affect the result of the lifetime calculation.  We therefore only varied
those parameters which have a substantial effect on the lifetime
calculation.  Maximum and minimum values for parameters that we did not
vary are given by ellipses.}\label{tbl:kda_params}

\end{deluxetable*}

\subsection{Treatment of the ICM}
\label{subsec:model_icm}

Apart from an assumption of self-similar growth, the KDA model also assumes
that the ambient medium is described by a power-law profile of the form
\begin{equation}
\label{eq:density}
\rho_x = \rho \left( \frac{r}{a} \right)^{-\beta}.
\end{equation}
where $a$ is the scale length of the distribution and $\rho$ is the density
at the scale length.  As every other equation in the KDA model involving
$\rho$ and $a$ incorporates this one, these two parameters are not
independent, but instead enter into the model only in their combined form,
$\rho a^{\beta}$ ($\beta$, however, enters the model in separate equations
and so is independent from $a$ and $\rho$).  \citet{kaiser07} term the
product $\rho a^{\beta}$ the ``density parameter'' and we adopt that
nomenclature in this paper.  

The KDA model uses this simplified profile to make the calculation of more
complicated quantities feasible, although at the expense of some
shortcomings in the model.  For example, while the KDA model adopts a
density distribution for the ambient medium, there is no model for the
pressure or the temperature.  As a result, the KDA model tacitly assumes
that the jet cocoon remains overpressured throughout its entire lifetime.
Indeed, the KDA model requires that the lobe drive a strong shock into the
external medium.  \citet{kaiser07} point out that the pressure in the
cocoon of an FR~II source decreases over time, and argue that if the cocoon
pressure falls below the ambient pressure at any point along the surface of
the cocoon, Kelvin-Helmholtz instabilities will mix cooler gas from the
external medium into the cocoon.  This process would disrupt the jet and
transform it into an FR~I.  If the pressure profile can be described by a
flat core with a power-law tail, then low-power jets will become
underpressured and transform into FR~Is before escaping from the core
region; however, more powerful jets will escape the core region and will
remain overpressured (and hence FR~IIs) for their entire lifetimes.  With
this model, \citet{kaiser07} present an order-of-magnitude calculation and
find that the critical power is $3 \times 10^{37}$ W.  

Given the lengths of the FR~IIs in our sample, the KDA model predicts that
even FR~IIs with this minimum power are much more luminous than the sources
in our sample.  As described next in \S\ref{subsec:q_measurement}, to
reproduce jets with roughly the same lengths and luminosities as the jets
in our sample, the KDA model requires that the typical jet power be over an
order of magnitude smaller than the minimum FR~II jet power power found by
\citet{kaiser07}.  This would imply that the jets are underpressured prior
to leaving the flat core of the ICM.  There are two problems with this
interpretation, however.  The first is that a detailed examination of the
pressure profiles of galaxy groups and clusters reveals that, while complex
and varied in shape, they generally have cusps at their centers rather than
flat cores \citep[e.g.,][]{vikhlinin06}.  This suggests that many
underpowered jets would start their lives as FR~Is rather than later
disrupting from FR~IIs.  Because the ICM density profiles measured by
\citet{vikhlinin06} can vary substantially from one group or cluster to the
next, there is likely no single, universal critical power.  Furthermore,
because the ICM profile can exhibit plateaus and bumps, it is difficult to
predict whether an FR~II will ever become underpressured, and if so, at
what radius, without knowing the detailed ICM profile of a group or
cluster.  The second problem with the interpretation of the jets in our
sample as underpressured is that the KDA model calculates the pressure at
the head of the jet and assumes the pressure in the cocoon to be uniform.
Disruption of the jet is likely to occur near the base of the jet, however,
where the ICM pressure is greatest.  Since the jet expansion speed can be
comparable to, or even larger than, the sound speed within the cocoon, the
pressure within the cocoon at the head of the jet will not necessarily be a
good proxy for the pressure within the cocoon near the center of the FR~II
source, particularly towards the end of the jet's life.  Indeed,
hydrodynamical simulations suggest that the pressure within the cocoon
varies along and across the axis of an FR~II jet and that the cocoon can
become underpressured with respect to the ambient medium without disrupting
the jet \citep{carvalho02b}.  Based on these considerations and our
observations, we conclude that the minimum FR~II jet power derived by
\citet{kaiser07} is too high by at least an order of magnitude.

The density profile of the ICM in galaxy clusters is difficult to measure
and the core density and scale radius can vary by several orders of
magnitude \citep{mohr99, mulchaey00, vikhlinin06, freeland11}.
\citet{cotter96b} and \citet{kaiser97b} found that the power-law slope of
the ICM density profiles in galaxy clusters hosting FR~IIs are typically
consistent with a value of $\beta = 1.9$.  We adopt this slope throughout
this paper.  Larger values of $\beta$ cannot support FR~IIs \citep{falle91}
and smaller values result in a smaller variation in the density parameter
in different ICMs, thereby making our lifetime results less sensitive to
the ICM properties.  \citet{bird08} demonstrated in their appendix that the
lifetimes of FR~IIs depend upon the cube root of the density parameter, so
a three order-of-magnitude variation in density parameter manifests itself
as an order-of-magnitude uncertainty in the lifetime calculation of the
jet.  Although this dependence is weak, we note that quantities which
depend on the volume of the lobe (e.g., the energy contained in the lobe)
depend linearly on the density parameter.

Insufficient X-ray data exist to measure the ICM density profiles for these
clusters.  Instead we calculate the FR~II lifetime separately for five
different ICM density estimates drawn from the literature and then use
optical richness as a proxy for ICM density in \S\ref{sec:correlations}.
To facilitate reference to each of the density estimates we label each with
the symbol $\chi_i$.  These five density estimates are listed in full in
Table \ref{tbl:densities} with references, but we describe them in order of
increasing density parameter here as follows.  $\chi_1$ is the density
assumed by \citet{blundell99}, which was taken from \citet{garrington91}
and assumed to be typical of a poor group.  $\chi_2$ is the default density
assumed by \citet{kaiser97b} and was taken from \citet{canizares87}, who
also found it to be typical for poor groups.  $\chi_3$ was found by
\citet{jetha07} to be typical for moderately sized groups and poor clusters
(i.e.~a typical richness of 9 -- 12).  $\chi_4$ and $\chi_5$ were measured
by \citet{jones84}, who found them to be typical for moderately sized and
massive clusters, respectively.  These last two models were used by
\citet{kaiser99a}. These five estimates span three orders of magnitude and
the full range of plausible density parameters for the host clusters of the
FR~IIs in our sample.  We expect, however, that the \citet{jetha07} density
estimate most closely approximates the ICM density of most of the host
clusters in our sample since the clusters in our sample have a median
richness which is comparable to the median richness of the \citet{jetha07}
sample.\footnote{Although \citet{jetha07} do not cite richness estimates,
compiling richnesses for clusters in their sample from various sources in
the literature yields a median richness of nine.}  We therefore take the
$\chi_3$ density model to be the default density model throughout this
paper. 

\begin{deluxetable*}{lccccc}
\tablewidth{0pt}
\tablecolumns{6}
\tablecaption{ICM Density Profiles}
\tablehead{
\colhead{Model} &
\colhead{$a$} &
\colhead{$\rho$} &
\colhead{$\rho a^{\beta}$} &
\colhead{Environment} &
\colhead{Reference} \\
\colhead{} &
\colhead{(kpc)} &
\colhead{(kg m$^{-3}$)} &
\colhead{(kpc$^{1.9}$ kg m$^{-3}$)} &
\colhead{} &
\colhead{} \\
}

\startdata
$\chi_1$ & 10 & $1.7 \times 10^{-23}$ & $1.3 \times 10^{-21}$ & Field \& poor groups & \citet{blundell99, garrington91} \\
$\chi_2$ & 2 & $7.2 \times 10^{-22}$ & $2.7 \times 10^{-21}$ & Field \& poor groups & \citet{kaiser97b, canizares87} \\
$\chi_3$ & 391 & $7.2 \times 10^{-26}$ & $6.1 \times 10^{-21}$ & Large groups \& poor clusters & \citet{jetha07} \\
$\chi_4$ & 30 & $5.0 \times 10^{-23}$ & $3.2 \times 10^{-20}$ & Moderate clusters & \citet{kaiser99a, jones84} \\
$\chi_5$ & 260 & $5.9 \times 10^{-24}$ & $2.3 \times 10^{-19}$ & Large clusters & \citet{kaiser99a, jones84} 
\enddata

\tablecomments{Density profiles given in terms of increasing density
parameter, $a \rho^{\beta}$.  The density and scale length are not
independent parameters in the KDA model, but only appear in their combined
form $a \rho^{\beta}$.}\label{tbl:densities}

\end{deluxetable*}

\subsection{The Jet Power Distribution}
\label{subsec:q_measurement}

In addition to an ICM model, the KDA model requires a jet power to
calculate the length and luminosity as a function of age.  To model the
observed length distribution, we therefore require a plausible distribution
of jet powers.  \citet{bird08} used two power-law distributions, one from
\citet{blundell99} and one derived from \citet{sadler02}.  Although the two
distributions have very different slopes (0.62 and 2.6), they have similar
median powers.  Because the FR~II length distribution depends primarily on
the median power of the FR~IIs, not powers at the tails of the
distribution, \citet{bird08} found similar results with both distributions.
We find both of these distributions to be unsuitable, however.  If we use
the KDA model to model a jet at the median power of these distributions
($\sim$10$^{38}$ W) and with a length typical for the jets in our sample
($\sim$100 kpc), the predicted radio luminosity is two orders of magnitude
larger than the typical radio luminosity of the FR~IIs in our sample.

As \citet{bird08} showed, lifetime estimates scale as the cube root of the
median jet power.  Nevertheless, an error in the power distribution by over
an order of magnitude can produce errors in the lifetime calculations by a
factor of at least three.  To accurately determine the jet lifetime, we
therefore need a reasonable estimate of the jet powers in our sample.

\begin{figure}
\centering
\includegraphics[width=8cm]{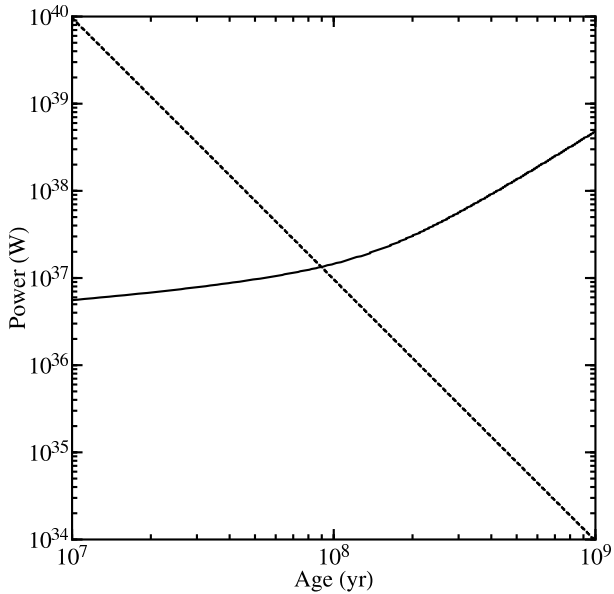}

\caption{A graphical explanation of the procedure to find the age and power
of an FR~II source given its length and luminosity. The solid line is the
locus of points in age-power space consistent with the jet's observed
luminosity, in this case $1.3 \times 10^{25}$ W Hz$^{-1}$.  The dashed line
is the locus of points consistent with the jet's observed length, in this
case 98.1 kpc.  The intersection of the two lines is the only age and power
consistent with the observed length and luminosity of the jet within the
KDA model.  The data shown in this plot are for a typical jet in our sample
for an axial ratio of $A = 4$ and an ICM with $\rho = 7.2 \times 10^{-26}$
kg m$^{-3}$ and $a = 391$ kpc.  Because the observed length of an FR~II is
shorter than its true physical length due to projection effects, we make an
average correction to the observed jet length before estimating its age and
power.\label{fig:const_lum_len}}

\end{figure}

Since the KDA model is a one-to-one function of age and power to length and
radio luminosity over the range of physically-motivated choices of the
input and auxiliary parameters, it is possible to numerically invert it to
produce an age and power given a length and luminosity.  For any given jet
in our sample, we compute the set of powers which reproduces the observed
length over a range of ages from $10^6$ yr to about $5 \times 10^9$ yr (at
which point numerical calculations begin to stop
converging\footnote{Specifically, for sufficiently extreme parameters,
Equation (A26) of \citet{kaiser07} fails to have a solution.}).  This upper
bound is well above numerous other constraints on the lifetime of AGN, and
FR~IIs specifically \citep[e.g.,][]{martini04, o'dea09}.  Because the
observed projected length is only a lower bound on the true physical length
of the FR~II, we make an average correction of a factor of $4/\pi$ to the
length to account for this projection effect.  We similarly compute the set
of powers which reproduces the observed luminosity over the same range of
ages.  The intersection of the loci of the two sets is then an estimate of
the age and power of the FR~II.  Figure~\ref{fig:const_lum_len} illustrates
the locus of constant length and the locus of constant luminosity for a
typical FR~II in our sample.

\begin{figure}
\centering
\includegraphics[width=8cm]{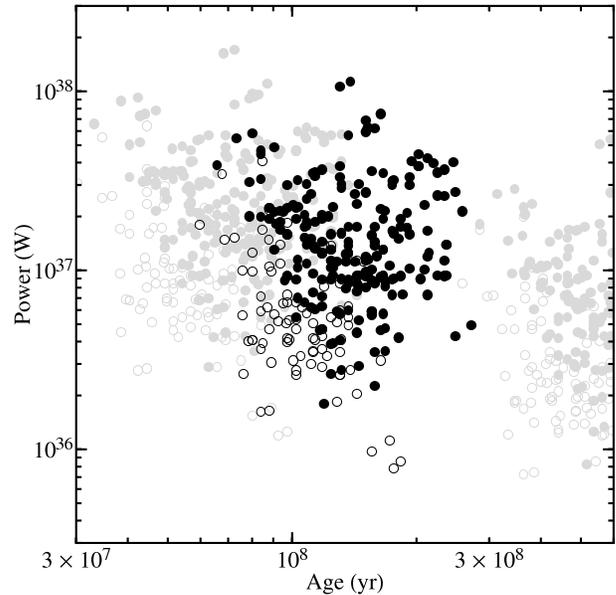}

\caption{The distribution of the full sample (filled and open circles) and
the volume-limited sample (filled circles only) in estimated jet power and
age assuming an axial ratio of 4 and the $\chi_3$ density model suitable
for typical clusters (black points), the $\chi_1$ density model suitable
for galaxy groups (left group of light gray points), and the $\chi_5$
density model suitable for the richest clusters (right group of light gray
points).  (See Table \ref{tbl:densities} for definitions of the density
models.)  Because the age and power estimates indicated here are determined
simultaneously, errors in the power estimate can induce errors in the age
estimate, and vice versa.  Furthermore, although an average correction is
made for projection effects, the age and power estimates in this figure do
not reflect any distribution in projection angle.  The age estimates in
this figure are therefore only approximate.  We use the length distribution
to constrain the age more precisely because it is more strongly dependent
on age than luminosity.  Because the uncertainty in our estimates is
dominated by systematic uncertainties in the ICM density and axial ratios
of the FR~IIs, we do not show error bars in this plot.  Systematic
uncertainties are on the order of a factor of two in both age and jet power
based on variation in these estimates over a reasonable range of ICM
densities and axial ratios.\label{fig:q_age}}

\end{figure}

We perform this process for every jet in our sample with each of the five
ICM models.  The median powers and ages are presented in Table
\ref{tbl:estimates}, and the distribution of the sample in power-age space
for the \citet{jetha07} ICM density estimate is displayed in
Figure~\ref{fig:q_age}.  These median powers range from $4 \times 10^{36}$
W to $10^{37}$ W, lower than the median jet power assumed in \citet{bird08}
by over an order of magnitude.  These jet powers are also lower than the
minimum FR~II jet power derived in \citet{kaiser07}; we provide a plausible
explanation for why their calculation is not inconsistent with our results
in \S\ref{subsec:model_icm}.  Our results are consistent with other,
simpler methods of estimating the jet power \citep[e.g.,][]{punsly05}. 

\begin{deluxetable}{cccc}
\tablewidth{0pt}
\tablecolumns{4}
\tablecaption{Estimated Median Jet Powers and Ages}
\tablehead{
\colhead{ICM Density} &
\colhead{$A$} &
\colhead{$Q_{\textrm{med}}$} &
\colhead{$t$} \\
\colhead{Model} &
\colhead{} &
\colhead{($10^{35}$ W)} &
\colhead{($10^7$ yr)} \\
\colhead{(1)} &
\colhead{(2)} &
\colhead{(3)} &
\colhead{(4)}
}

\startdata
$\chi_1$ & 2 & 211.3 & 9.8 \\
$\chi_1$ & 3 & 165.6 & 7.7 \\
$\chi_1$ & 4 & 145.0 & 6.6 \\
$\chi_1$ & 6 & 124.7 & 5.5 \\
$\chi_1$ & 8 & 113.9 & 4.8 \\
$\chi_1$ & 12 & 102.0 & 4.0 \\
$\chi_1$ & 16 & 95.1 & 3.5 \\
$\chi_2$ & 2 & 174.6 & 13.2 \\
$\chi_2$ & 3 & 134.6 & 10.4 \\
$\chi_2$ & 4 & 116.5 & 9.1 \\
$\chi_2$ & 6 & 99.0 & 7.3 \\
$\chi_2$ & 8 & 89.8 & 6.6 \\
$\chi_2$ & 12 & 79.7 & 5.4 \\
$\chi_2$ & 16 & 74.0 & 4.8 \\
$\chi_3$ & 2 & 142.9 & 18.5 \\
$\chi_3$ & 3 & 108.8 & 14.9 \\
$\boldsymbol{\chi}_\mathbf{3}$ & \textbf{4} & \textbf{93.3} & \textbf{12.8} \\
$\chi_3$ & 6 & 78.5 & 10.6 \\
$\chi_3$ & 8 & 70.7 & 9.3 \\
$\chi_3$ & 12 & 62.3 & 7.7 \\
$\chi_3$ & 16 & 57.5 & 6.9 \\
$\chi_4$ & 2 & 99.4 & 36.3 \\
$\chi_4$ & 3 & 75.0 & 29.1 \\
$\chi_4$ & 4 & 63.9 & 25.4 \\
$\chi_4$ & 6 & 53.1 & 20.8 \\
$\chi_4$ & 8 & 47.6 & 18.3 \\
$\chi_4$ & 12 & 41.6 & 15.2 \\
$\chi_4$ & 16 & 38.3 & 13.7 \\
$\chi_5$ & 2 & 68.4 & 79.2 \\
$\chi_5$ & 3 & 51.9 & 62.7 \\
$\chi_5$ & 4 & 44.5 & 54.6 \\
$\chi_5$ & 6 & 37.3 & 45.7 \\
$\chi_5$ & 8 & 33.6 & 39.7 \\
$\chi_5$ & 12 & 29.7 & 32.9 \\
$\chi_5$ & 16 & 27.5 & 29.2 \\
\enddata

\tablecomments{The median powers and ages of the FR~IIs in the full sample
estimated using the method described in \S\ref{subsec:q_measurement}.  Our
default choice of density model and axial ratio is shown in bold.  (1): The
density model (defined in Table \ref{tbl:densities}).  (2): The axial ratio
(defined in \S\ref{subsec:axial_ratio_dependence}).  (3): The estimated
jet power.  (4): The estimated current age.  Due to selection effects there
is no simple way to convert the median current age to a lifetime
estimate.}\label{tbl:estimates}

\end{deluxetable}

\begin{figure}
\centering
\includegraphics[width=8cm]{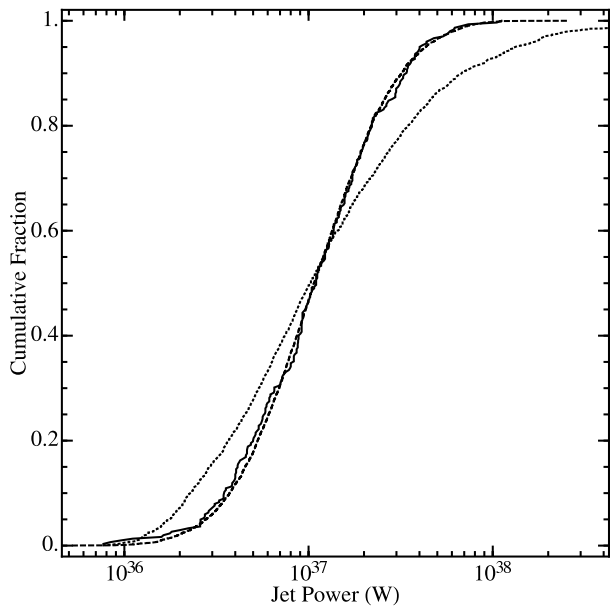}

\caption{The cumulative distribution of jet power for FR~IIs in our sample
(solid line), the best-fitting power law distribution (dotted line), and
the best-fitting log-normal distribution (dashed line).  These
distributions are the observed distributions from mock catalogs which take
selection effects into account to make them directly comparable to the
observed sample.  The power law distribution shown here makes the
implausible requirement that $\sim$60\% of BCGs in the MaxBCG harbor FR
IIs, but only $\sim$7\% are detected.  Although better-fitting power law
distributions exist, they require even smaller detection fractions; even if
all BCGs in the MaxBCG catalog harbored FR~IIs, these distributions predict
FR~II sample sizes much less than the 151 we observe.  The log-normal
distribution, by contrast, predicts a detection fraction of $\sim$60\%.}
\label{fig:lognormalq}

\end{figure}

The resulting distribution of FR~II jet powers in our sample is not well
described by a truncated power law.  To account for selection effects, we
generate a set of mock catalogs (described in further detail in
\S\ref{subsec:mock_generation}) consisting of $10^6$ FR~IIs given random
ages, redshifts, and orientations.  If we assume that the jet powers are
distributed as a truncated power law with a minimum power
$Q_{\textrm{min}}$ and a slope $\alpha$, a maximum likelihood analysis
(described in further detail in \S\ref{subsec:maxlik}, where it is applied
to the length distribution and lifetime measurement) applied to the
resulting power distributions indicates that the best fit is
$Q_{\textrm{min}} = 7.6 \times 10^{35}$ W, $\alpha = 2.2$ for the
\citet{jetha07} ICM model and an axial ratio of 4.  Although better fits
exist in principle, the fraction of detectable FR~IIs in them is so low
that our sample size would be much less than 151 even if every BCG in the
MaxBCG catalog hosted an FR~II.  Even in this best fit, a sample size of
151 FR~IIs can only be recovered if $\sim$50\% of the BCGs in the MaxBCG
catalog host FR~IIs. We find that the power distribution is much better
described by a log-normal distribution.  A similar analysis shows that the
best fit has a median power of $1.1 \times 10^{37}$ W and a coefficient of
variation of 2.2 under the same density and axial ratio assumptions.  (We
perform this fit for every choice of ICM density and axial ratio we examine
in this work.  See \S\ref{subsec:axial_ratio_dependence} and
\S\ref{subsec:icm_dependence} for these choices.)  A log-normal
distribution not only fits the observed jet power distribution better than
a power law, but it also does not require a very low detection probability,
and hence does not require that a large fraction of BCGs harbor FR~IIs
($\sim$2\% for a log-normal distribution vs.~$\sim$60\% for a power law).
The cumulative distribution of jet powers is presented in Figure
\ref{fig:lognormalq} along with the best-fitting power law and log-normal
distributions.  

The true distribution is likely more complicated (and possibly an
approximate power law) than log-normal, but only appears log-normal as a
result of selection effects.  Although the mock catalogs account for any
biases due to our selection algorithm, they do not account for selection
effects imposed by the MaxBCG catalog.  Specifically, the distribution of
BCG stellar luminosities in the MaxBCG catalog can be much better described
by a log-normal distribution than a power law.  Since BCG stellar
luminosities are correlated with FR~II jet powers (as we show in
\S\ref{subsubsec:Q_lum_correlation}), it is not surprising that our
observed distribution of jet powers resembles a log-normal distribution.
The log-normal distribution in BCG luminosities may be the result of the
steep galaxy luminosity function at high luminosities and the minimum
richness cutoff of the sample.  A study of jet powers in lower-luminosity
galaxies and field galaxies could determine if the intrinsic jet power
distribution is log-normal for all galaxies.

While the method described in this section produces estimates of the ages
of the FR~IIs, we do not expect them to be particularly precise.  This is
because the estimates of the lengths and ages of the FR~IIs place equal
weight on the equation for luminosity evolution and the equation for length
evolution.  The length evolution of the FR~II is relatively straightforward
to model, because it only relies on dimensional analysis, and hence is
consistent across many models.  The luminosity evolution, on the other
hand, is much more difficult to model and so has greater uncertainty.  Even
if the equation for luminosity evolution were known perfectly, there is
also a greater uncertainty in the flux measurement than in the length
measurement.  This is because the sources are extended and therefore flux
from diffuse components can be lost and flux from spatially coincident
sources can be added.  We therefore only use this process to estimate the
jet power to within a factor of a few and then use more precise methods to
measure the lifetime.  Because the lifetime calculation depends only on the
cube root of the jet power, the uncertainty in the power estimate will not
introduce an error in the lifetime calculation of more than a factor of
two.

\subsection{The $Q-P_{1.4}$ Relation}
\label{subsec:pjet_lum}

Several authors have sought a simple power-law relationship between the jet
power and radio luminosity \citep{willott99, punsly05, birzan08,
cavagnolo10}.  Although the jet power is the fundamental physical parameter
of interest in radio galaxies, it is not easily measurable.  A simple,
approximate conversion between jet power and the observed radio luminosity
would be useful to estimate jet powers in large samples of radio galaxies.

\begin{figure}
\centering
\includegraphics[width=8cm]{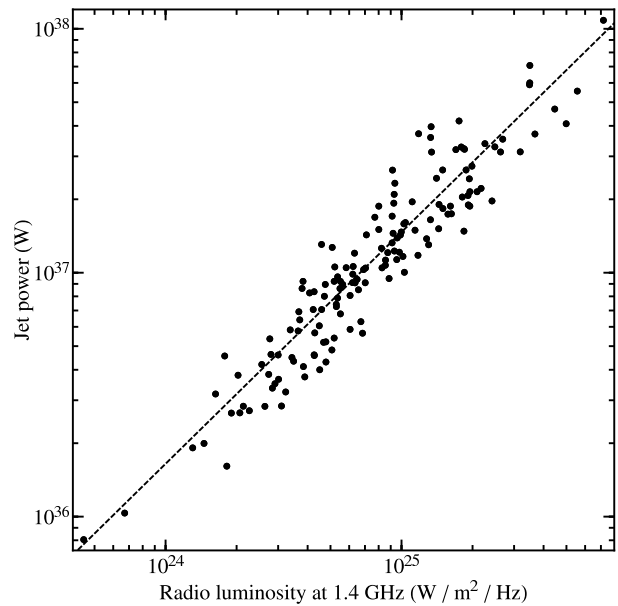}

\caption{The best-fit power law relation between radio luminosity at 1.4
GHz and the jet power assuming the $\chi_3$ density model (defined in Table
\ref{tbl:densities}) and an axial ratio of 4.  Changes to the ICM density
and axial ratio change the overall normalization but do not have a large
effect on the exponent.  The method for calculating the jet powers is
described in \S\ref{subsec:q_measurement}.  We emphasize that the best fit
relation is heavily model-dependent due to our use of the KDA model to
convert from jet length and radio luminosity to the jet power.  This
relation is therefore best interpreted more as a prediction of the KDA
model than a direct measurement of the $Q - P_{1.4}$ relation.}
\label{fig:q_radiolum}

\end{figure}

We fit the powers estimated in \S\ref{subsec:q_measurement} to a power-law
in 1.4 GHz radio luminosity.  For the \citet{jetha07} ICM density model and
an axial ratio of 4, we find the best fit to be
\begin{equation}
\label{eq:pjet_lum}
\log Q = 0.95(\pm 0.03) \log P_{1.4} + 13.4(\pm 1.1)
\end{equation}
where both $Q$ and $P_{1.4}$ are in Watts and the quoted errors are purely
statistical.  Variation of the ICM density and axial ratio changes the
normalization, but leaves the exponent of the power law unchanged to within
the quoted error.  This fit is shown in Figure \ref{fig:q_radiolum}.  The
scatter in this relation is largely due to the fact that variations in
length (and hence age) are ignored.

This relation is steeper than the relations found by \citet{birzan08} and
\citet{cavagnolo10}, who find exponents of $0.35 \pm 0.07$ and $0.75 \pm
0.14$, respectively.  Our relation is, however, consistent with the
relation between $Q$ and the radio luminosity at 151 MHz derived by
\citet{willott99} and \citet{punsly05}, who found an exponent of 0.86.  

We emphasize that our fit is heavily model-dependent because our conversion
from jet length and radio luminosity to jet power is derived from the KDA
model.  While the KDA model likely provides estimates of jet powers which
are correct to within a factor of a few, any errors in the relationship
between jet power and radio luminosity derived in the model will be
reflected in the exponent of our fit.  The scatter in this relationship
will also be influenced by the intrinsic jet power and age distribution of
the sample.  The scatter for our sample may be relatively modest because
the jet power distribution follows a log-normal distribution, which may not
be the case for all FR~II samples due to the selection of the input cluster
catalog (see \S\ref{subsec:cluster_catalog}).

\subsection{Mock Catalog Generation}
\label{subsec:mock_generation}

We compare the observed length distribution of FR~IIs to mock catalogs to
infer their intrinsic properties.  These mock catalogs are based on large,
simulated datasets that sample a suite of KDA model parameters (e.g., the
ICM density and axial ratio) and a range of maximum ages.  For any given
choice of KDA model parameters, we create 31 mock catalogs with maximum
ages ranging from $1.1 \times 10^7$ to $1.2 \times 10^9$ yr and distributed
evenly logarithmically.  Roughly $1.3 \times 10^6$ jets are then created by
randomly choosing an age, power, and redshift for each jet.  The age is
sampled from a uniform distribution ranging from 0 to the maximum age
chosen for that particular catalog and the jet power is sampled from the
log-normal distribution derived in \S\ref{subsec:q_measurement}.  To
account for selection effects, we decrease the median power by 15\%, an
empirically determined amount for which the jet power distribution of the
mock catalogs best matches the observed jet power distribution.  The
redshift of the simulated jet is then sampled from the redshift
distribution of the entire MaxBCG catalog.  (The redshift distribution is
necessary when calculating the jet luminosities due to the effect of the
CMB energy density on synchrotron losses.)  We assume that the birth rate
of FR~IIs is constant across the redshift range of the MaxBCG catalog.  The
support for this assumption is discussed in \S\ref{subsec:friifrac}.

Once a jet has been created, we calculate its $K$-corrected luminosity and
randomly orient it on the sky to calculate its projected length.  We then
apply the selection criteria described in \S\ref{subsec:crosscorr} to
create a mock catalog for that parameter set.  

\section{The FR~II Lifetime}
\label{sec:frii_lifetime}

\subsection{Maximum Likelihood Fitting}
\label{subsec:maxlik}

Our approach to estimate the maximum lifetime for FR~IIs is to compare the
projected length distribution of the mock catalogs to the observed
distribution.  We fit the observed distribution to the mock catalogs with
the maximum likelihood method.  The result is the best-fit maximum age for
a given set of model parameters.

The likelihood of any maximum age is determined by constructing a
probability distribution function (PDF) of the length distribution for a mock
catalog with 1-kpc-wide bins.  If the number of jets in the mock
catalog that fall into the $i^{\textrm{th}}$ bin is $n_i$, the
probability that a jet will fall in that bin is $n_i / N$, where $N$ is the
total number of jets in the mock catalog, $\sum_i n_i$.  With a PDF, $f$,
so defined, the likelihood of any model is then
\begin{equation}
\label{eq:lik}
\mathcal{L} = \prod_j f(l_j),
\end{equation}
where $l_j$ is the projected length of the $j^{\textrm{th}}$ jet
in the observed sample.  For any given choice of parameters, we take the
best-fit lifetime to be the maximum age of the mock catalog which
maximizes the likelihood.  The error on the lifetime calculation is taken
to be the FWHM of the likelihood distribution. 

\begin{figure}
\centering
\includegraphics[width=8cm]{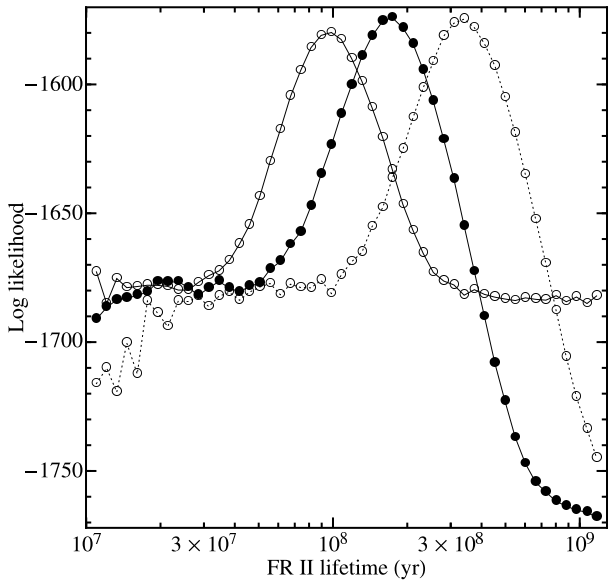}

\caption{Likelihood distributions of three models for a range of FR~II
lifetimes.  The peak of the likelihood distribution for a particular model
is the lifetime that best matches the observed data.  The solid points
represent the likelihood distribution for the default ICM density model
$\chi_3$ and an axial ratio of 4.  The open circles connected by a solid
line represent the default ICM density model $\chi_3$ and an axial ratio of
12.  The open circles connected by a dotted line represent the $\chi_4$ ICM
density model and an axial ratio of 4.  See Table \ref{tbl:densities} for a
description of the ICM density models.\label{fig:liks}}

\end{figure}

\begin{figure*}
\centering
\includegraphics[width=16cm]{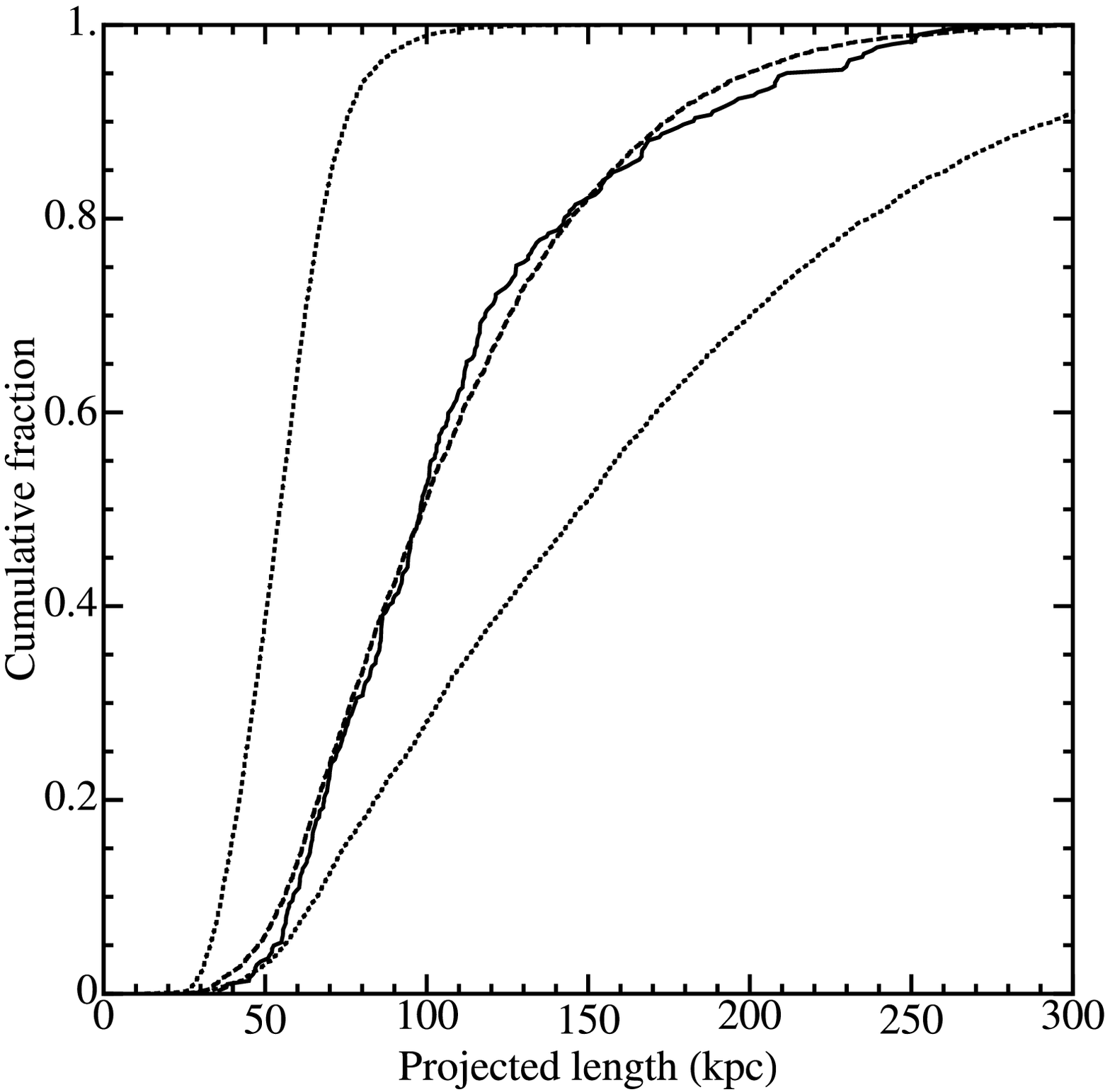}

\caption{A comparison of the cumulative projected length distribution of
the sample (solid line) to the cumulative length distributions of a model
for three different lifetimes (dashed and dotted lines) assuming the
default ICM density $\chi_3$ (see Table \ref{tbl:densities}) and an axial
ratio of 4.  The dashed line shows the FR~II lifetime from the model with
the maximum likelihood, which in this case was $1.9 \times 10^8$ yr.  The
left and right dotted lines represent FR~II lifetimes of $6.1 \times 10^7$
yr and $3.4 \times 10^8$ yr, respectively. \label{fig:mock_len_cumul}}

\end{figure*}

Figure~\ref{fig:liks} shows likelihood distributions for several models.
The length distribution of the mock catalog for a sample model at three
ages is displayed with the observed length distribution in
Figure~\ref{fig:mock_len_cumul}.  Because the projected length distribution
is extremely sensitive to age, we can determine the FR~II lifetime with
relatively high precision.  Table \ref{tbl:lifetimes} presents the
lifetimes and their uncertainties for every model used.  We note that we
only measure the \emph{typical} FR~II lifetime and there is likely some
intrinsic dispersion which is larger than the statistical uncertainties
indicate.  We discuss any correlations between the typical FR~II lifetime
and various jet properties in \S\ref{subsubsec:lifetime_corr}.  The
best-fitting lifetime for our default ICM model is $1.9 \times 10^8$ yr. 

\begin{deluxetable}{ccc}
\tablewidth{0pt}
\tablecolumns{3}
\tablecaption{FR~II Lifetimes \& Statistical Errors}
\tablehead{
\colhead{ICM Density} &
\colhead{$A$} &
\colhead{$t_{\textrm{lifetime}}$} \\
\colhead{Model} &
\colhead{} &
\colhead{($10^7$ yr)} \\
\colhead{(1)} &
\colhead{(2)} &
\colhead{(3)} 
}

\startdata
$\chi_{1}$ & 2 & 15.5 $^{+0.3}_{-0.8}$ \\
$\chi_{1}$ & 3 & 11.9 $^{+0.4}_{-0.2}$ \\
$\chi_{1}$ & 4 & 10.0 $\pm 0.2$ \\
$\chi_{1}$ & 6 & 8.3 $\pm 0.2$ \\
$\chi_{1}$ & 8 & 7.1 $\pm 0.2$ \\
$\chi_{1}$ & 12 & 6.0 $\pm 0.4$ \\
$\chi_{1}$ & 16 & 5.2 $\pm 0.1$ \\
$\chi_{2}$ & 2 & 20.3 $\pm 0.4$ \\
$\chi_{2}$ & 3 & 16.2 $^{+0.6}_{-0.3}$ \\
$\chi_{2}$ & 4 & 14.1 $^{+0.6}_{-0.5}$ \\
$\chi_{2}$ & 6 & 11.7 $\pm 0.2$ \\
$\chi_{2}$ & 8 & 9.9 $\pm 0.2$ \\
$\chi_{2}$ & 12 & 8.0 $\pm 0.2$ \\
$\chi_{2}$ & 16 & 6.9 $\pm 0.4$ \\
$\chi_{3}$ & 2 & 28.1 $^{+0.2}_{-0.1}$ \\
$\chi_{3}$ & 3 & 23.2 $\pm 0.4$ \\
$\boldsymbol{\chi}_{\mathbf{3}}$ & \textbf{4} & \textbf{19.2} $\mathbf{\pm 0.4}$ \\
$\chi_{3}$ & 6 & 16.2 $\pm 0.3$ \\
$\chi_{3}$ & 8 & 14.4 $\pm 0.5$ \\
$\chi_{3}$ & 12 & 11.7 $^{+0.3}_{-0.5}$ \\
$\chi_{3}$ & 16 & 10.2 $\pm 0.4$ \\
$\chi_{4}$ & 2 & 55.9 $\pm 0.9$ \\
$\chi_{4}$ & 3 & 45.3 $\pm 1.8$ \\
$\chi_{4}$ & 4 & 38.1 $^{+1.5}_{-0.7}$ \\
$\chi_{4}$ & 6 & 31.5 $^{+1.8}_{-1.2}$ \\
$\chi_{4}$ & 8 & 27.6 $^{+1.0}_{-0.6}$ \\
$\chi_{4}$ & 12 & 23.6 $^{+0.5}_{-1.3}$ \\
$\chi_{4}$ & 16 & 19.9 $^{+0.8}_{-0.3}$ \\
$\chi_{5}$ & 2 & 122 $\pm 3$ \\
$\chi_{5}$ & 3 & 105 $^{+2}_{-4}$ \\
$\chi_{5}$ & 4 & 84.9 $\pm 1.7$ \\
$\chi_{5}$ & 6 & 68.9 $\pm 2.6$ \\
$\chi_{5}$ & 8 & 59.1 $^{+2.4}_{-1.2}$ \\
$\chi_{5}$ & 12 & 50.8 $^{+0.9}_{-2.9}$ \\
$\chi_{5}$ & 16 & 42.8 $^{+1.6}_{-0.8}$ \\
\enddata

\tablecomments{FR~II lifetimes with the maximum likelihood for every
density model and axial ratio.  The default choice of density and axial
ratio is in bold.  (1): ICM density model (defined in Table
\ref{tbl:densities}).  (2): Axial ratio (defined in
\S\ref{subsec:axial_ratio_dependence}).  (3): FR~II lifetime.  The quoted
uncertainties are purely statistical and are derived from the FWHM of the
likelihood distribution.}\label{tbl:lifetimes}

\end{deluxetable}

\subsection{Dependence on Axial Ratio}
\label{subsec:axial_ratio_dependence}

We define the axial ratio as the ratio of the length of the lobe to the
perpendicular distance between the edge of the lobe and its axis.  (Some
other studies define the axial ratio as the ratio between the length of the
lobe and the width of the lobe; our definition is twice this other
convention.)  Due to projection effects, background noise, and the fact
that the shapes of real radio lobes are more complicated than the simple
geometry of radio lobe models, unbiased and robust measurements of the
axial ratios of FR~II sources are difficult to obtain.  Nevertheless,
nearly all estimates fall between 2 and 16 \citep{leahy84, machalski04a}.
Some studies of extremely powerful jets have found even larger axial
ratios, but there appears to be a relationship between jet power and axial
ratio, with more powerful jets having unusually large axial ratios
\citep{leahy89}.  While it is possible to estimate the axial ratios of the
FR~IIs in our sample from the FIRST images, the width of the lobes are in
many cases comparable to the size of the FIRST PSF, so such estimates only
provide lower bounds on the axial ratios.  Since our sample does not
include atypically powerful jets, we simply examine axial ratios within the
entire plausible range of $ 2 \leq A \leq 16$ and note that our estimates
are most consistent with an axial ratio of 3 -- 4, but could be larger.
Higher-resolution data of some of the FR~IIs in our sample would be needed
to accurately measure their axial ratios.

To first order, one expects that as the axial ratio increases, the
lifetime should decrease.  At high axial ratios, the jet does not need to
expend as much energy inflating a fat cocoon around it and can therefore
drill through the ICM more efficiently; thus, a jet of given power with a
large axial ratio can reach the same length as a jet with a small axial
ratio in less time.  But this simple relationship is complicated by the
fact that the axial ratio changes the estimated power as well.  A jet with
a larger axial ratio requires less power to produce a given length in a
given time, so larger axial ratios will lead to lower power estimates.
This will partially counterbalance the increased age estimate and produce a
smaller axial ratio dependence on age than one would naively expect from
the jet length and age alone.  Nevertheless, because the lifetime is
relatively insensitive to the power, this correction does not dominate. 

\begin{figure}
\centering
\includegraphics[width=8cm]{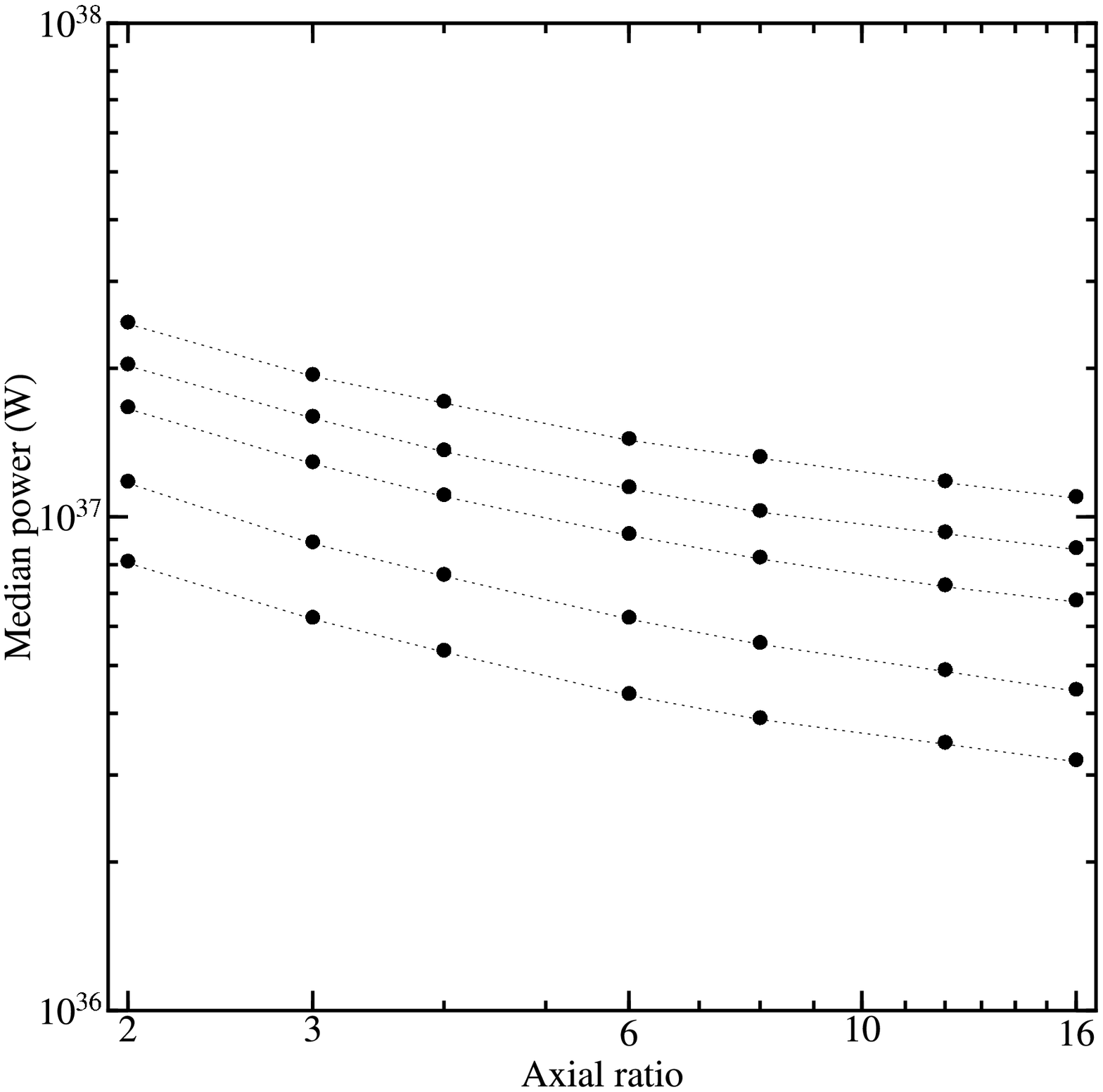}

\caption{The dependence of the axial ratio on the median inferred power of
the FR~II sample.  The points along each dotted line represent the median
powers inferred for a particular density model.  From bottom to top, the
density models used are $\chi_1$ -- $\chi_5$.  See Table
\ref{tbl:densities} for the details of each density model.
\label{fig:q_axial}}

\end{figure}

\begin{figure}
\centering
\includegraphics[width=8cm]{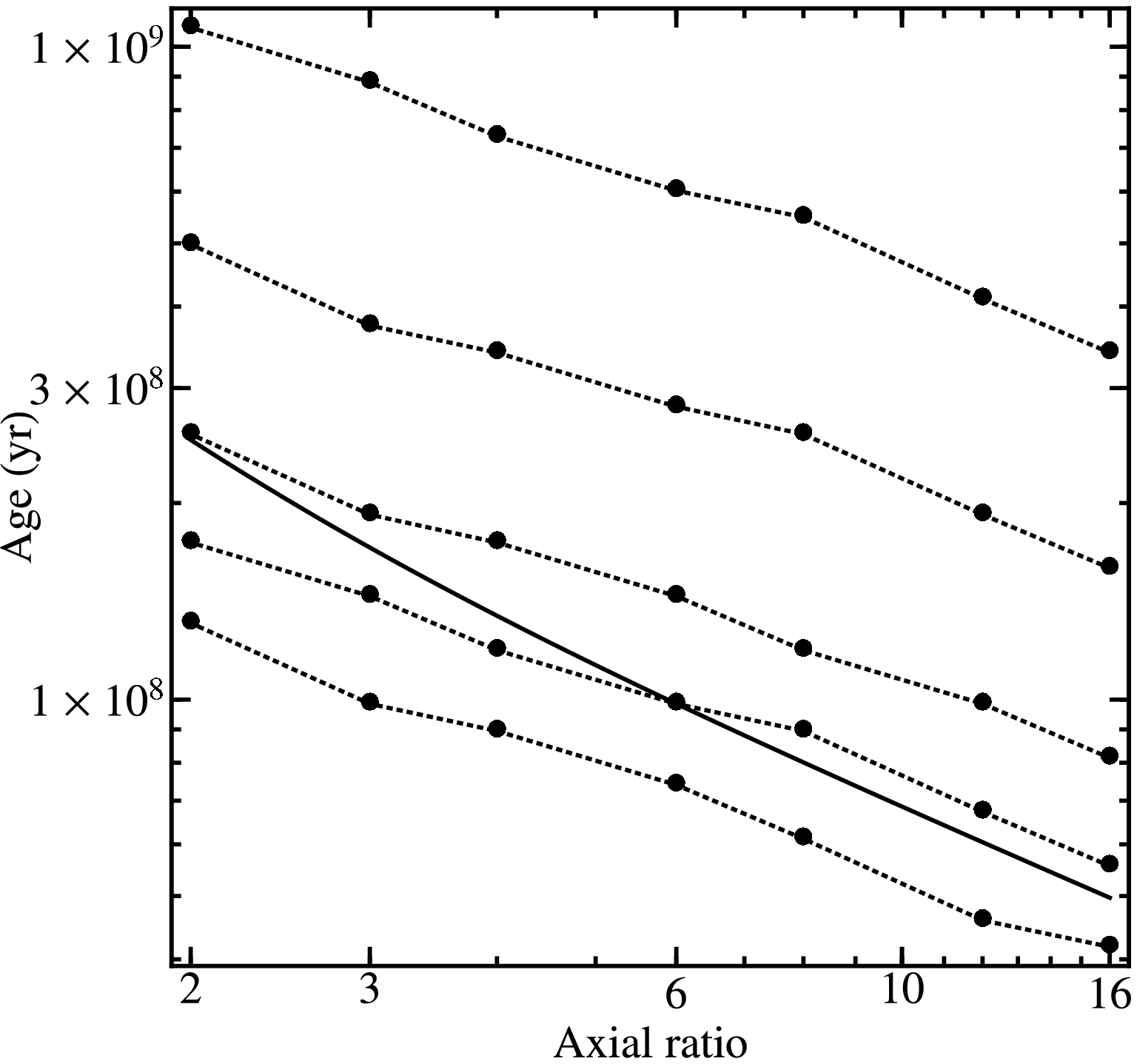}

\caption{The dependence of FR~II lifetime estimates on the assumed axial
ratio.  Each dotted line represents a set of lifetime estimates for a
single density.  Lower tracks assume lower densities.  The densities used
are listed in Table \ref{tbl:densities}.  The solid line represents the
axial ratio dependence one would expect if one only considered the effect
of axial ratio on the length of the jets as a function of time.  Because
the axial ratio also has an effect on our power estimates of the jet which
tends to counteract the effect of axial ratio on length evolution, the
final lifetime estimate is less sensitive to the axial ratio than found by
other studies \citep[e.g.,][]{bird08}.  \label{fig:t_axial}}

\end{figure}

\citet{bird08} found that the axial ratio has a stronger effect on FR~II
lifetime calculations than all other KDA model parameters.  To evaluate the
dependence of our FR~II lifetime measurements on axial ratio, we calculated
the lifetimes of jets in our sample for different axial ratios.  For any
particular choice of axial ratio, we used the KDA model to estimate the jet
powers, generated a set of mock catalogs across a range of ages, and fit
these to the observed length distribution with the maximum likelihood
method described in \S\ref{subsec:maxlik}.  The dependences of power and
lifetime measurements on the assumed axial ratio are shown in
Figures~\ref{fig:q_axial} and~\ref{fig:t_axial}.  The inferred power
decreases as the axial ratio increases, but the effect is only
approximately a factor of two decrease over a factor of eight increase in
axial ratio.  The age estimate similarly decreases as the axial ratio
increases, with the effect also being a roughly factor of two decrease over
a factor of eight increase in axial ratio.  This effect is even weaker when
the change in inferred jet power is taken into account.

\subsection{Dependence on the ICM Model}
\label{subsec:icm_dependence}

The model length distribution also depends on the ICM density parameter.
ICM density measurements of group and cluster samples can vary by several
orders of magnitude \citep{vikhlinin06, jetha07, freeland11}. Since the
density of the model ICM can potentially have a large effect on FR~II
lifetime estimates, we produced sets of mock catalogs spanning about two
orders of magnitude in density.  The density profiles we use and their
references are listed in Table \ref{tbl:densities}.  

At higher densities, a jet of a given power will require more time to grow
to a given length, so to first order, increased densities will lead to
increased lifetime estimates, approximately linearly.  However, increased
densities also increase the luminosity of a jet at a fixed power; hence
higher densities lead to lower power estimates.  This, in turn, will
result in yet longer lifetime estimates in second order. 

\begin{figure}
\centering
\includegraphics[width=8cm]{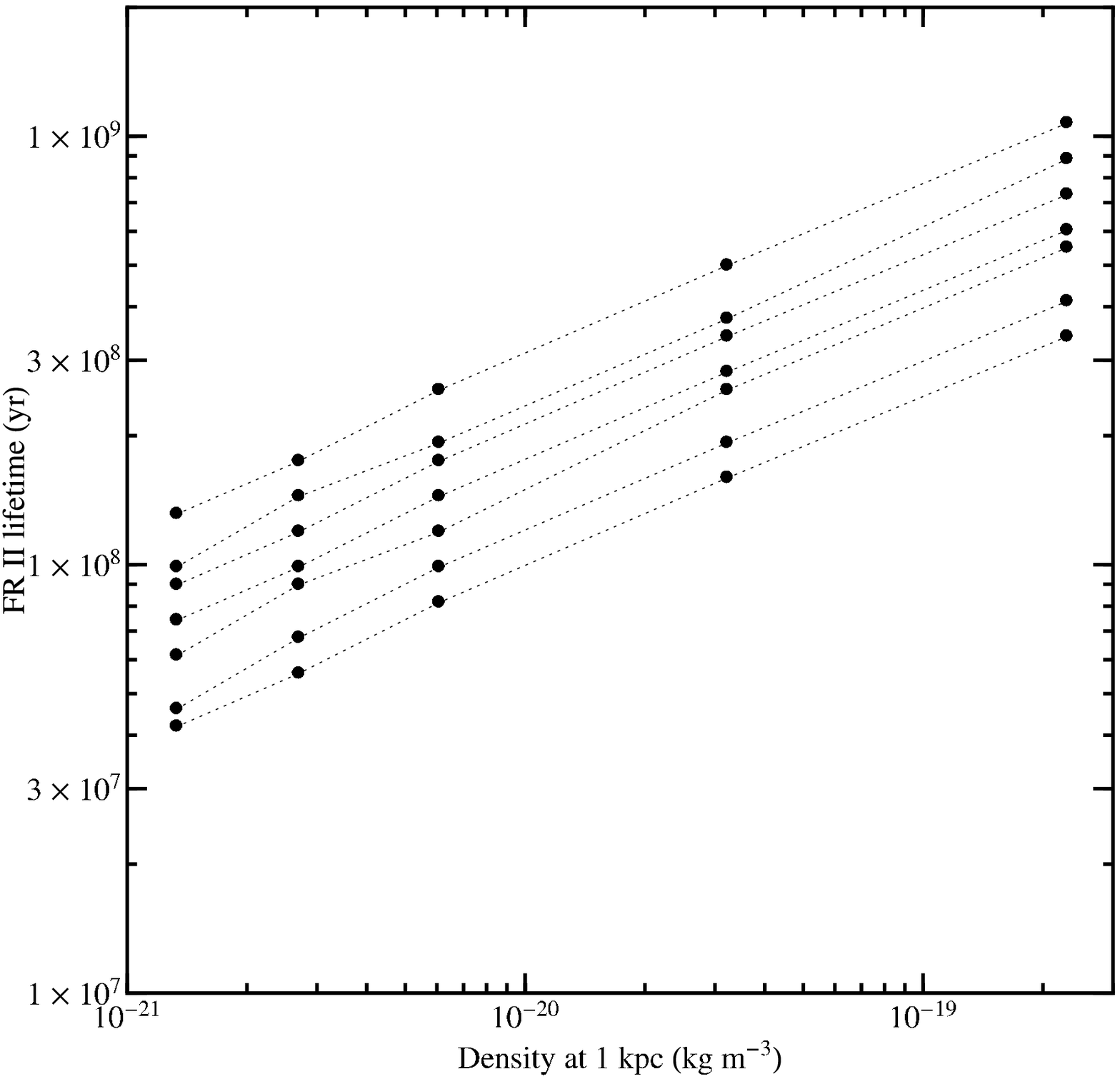}

\caption{The dependence of the FR~II lifetime calculation on the density of
the ICM.  In each case the density profile is described by a power law with
a slope of 1.9.  Each track represents a different choice of axial ratio;
lower tracks have larger axial ratios.  The values of the axial ratios are
the values of the points in Figures \ref{fig:q_axial} and
\ref{fig:t_axial}.\label{fig:t_density}}

\end{figure}

\begin{figure}
\centering
\includegraphics[width=8cm]{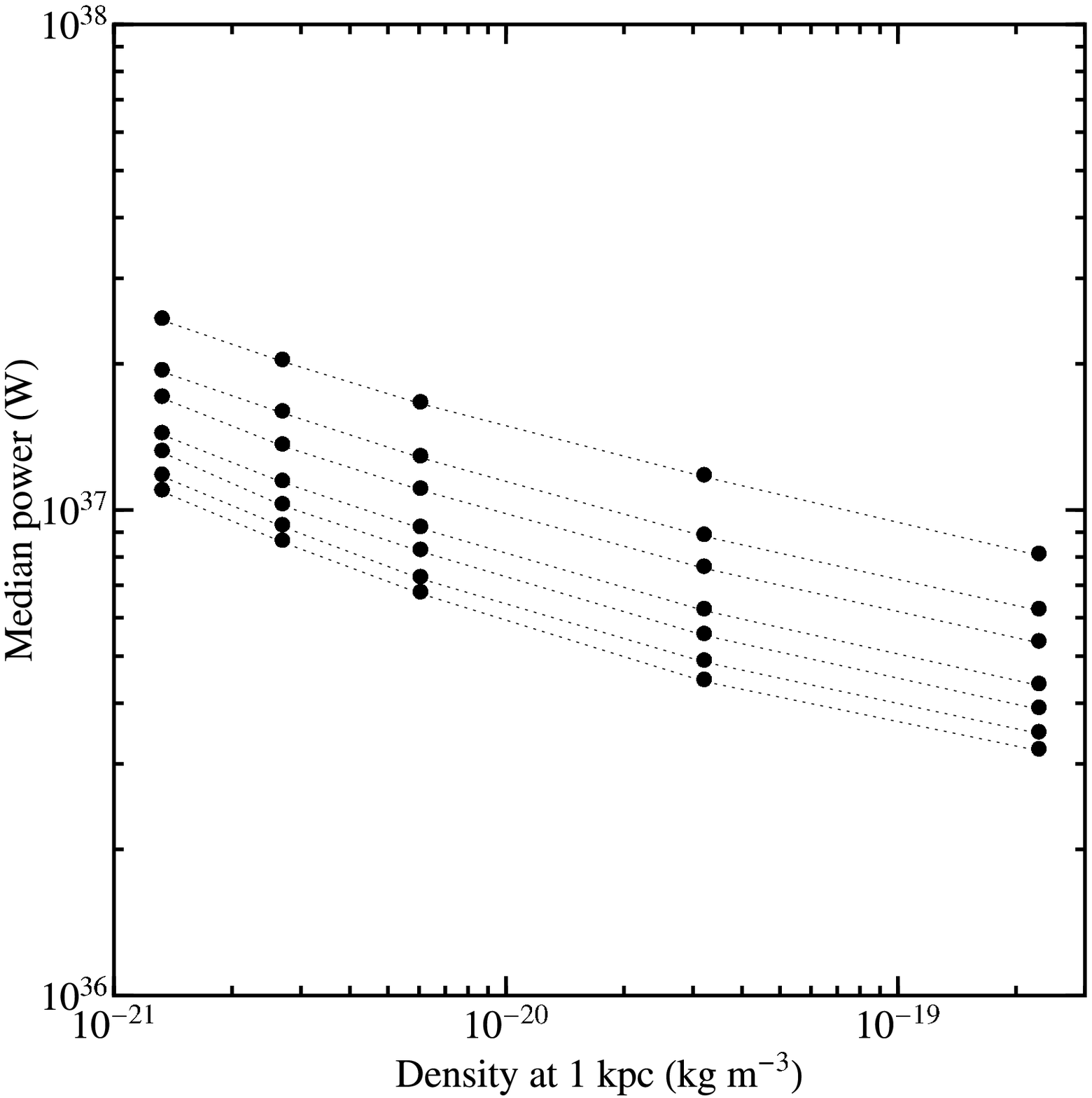}

\caption{The dependence of the estimated FR~II jet power on the density of
the ICM.  Each track represents a different choice of axial ratio; lower
tracks have larger axial ratios.  Refer to Figures \ref{fig:q_axial} and
\ref{fig:t_axial} for the axial ratio of each track.\label{fig:q_density}}

\end{figure}

For each ICM model, the length distribution of the mock catalogs was fit to
the observed length distribution of the flux-limited sample with the
maximum likelihood method described in \S\ref{subsec:maxlik}.  The
dependence of power and lifetime measurements on the assumed ICM model are
shown in Figures~\ref{fig:t_density} and~\ref{fig:q_density}.  These
figures indicate that larger densities yield larger lifetime estimates and
smaller power estimates, though in the case of the jet power estimates the
dependence appears to be more complicated in detail than a power law.

\subsection{The FR~II Duty Cycle}
\label{subsec:duty_cycle}

Because our mock catalogs take into account the biases induced by selection
effects on our sample, we use them to calculate a robust upper bound on the
duty cycle of FR~IIs in BCGs, where the FR~II duty cycle is the average FR
II lifetime divided by the fraction of time that the jet remains active.
Although it is not possible to directly measure the fraction of time that a
jet remains active, a lower limit on this fraction can be found in a
straightforward manner.  If we assume that every BCG in a cluster has a
radio jet at some point in its life, then the fraction of time that a jet
is active is the number of BCGs with radio jets divided by the total number
of BCGs.  Since it might be that many BCGs never have a radio jet, this
fraction is necessarily a lower limit.

The total number of BCGs in our sample is 13,823, the size of the MaxBCG
catalog of \citet{koester07b}.  The number of BCGs with FR~IIs is our
sample size of 151 FR~IIs plus some additional FR~IIs that fall below our
selection criteria.  To estimate the number of FR~IIs that are missed, we
calculate the fraction of simulated jets that pass all of the detection
thresholds in our mock catalogs.  The mock catalogs are populated with jets
resembling the underlying population of FR~IIs, so the fraction of jets
which pass our detection cuts is an estimate of the completeness of our
sample.  For the best-fit age and the default parameters, this completeness
fraction is 61.7\%.  The overwhelming majority of missed sources are too
short.  We divide the number of jets in our sample by this completeness
fraction and estimate that there are approximately 245 FR~IIs in these
BCGs.  The fraction of time that a jet is active is then 245/13,823, or
1.8\%.  This fraction is in agreement with the value found by
\citet{bird08} for galaxies in the group environment.  Combined with our
best estimate of the FR~II lifetime of $1.6 \times 10^8$ yr, this yields an
upper bound for the duty cycle of $9 \times 10^9$ yr.  As this upper bound
is comparable to a Hubble time, it implies that most BCGs undergo an FR~II
phase only once.  If a substantial fraction of BCGs are never FR~IIs (either
being always inactive or only appearing as FR~Is), then FR~IIs could be
episodic.

\section{The Relationship between FR~IIs and their Environments}
\label{sec:correlations}

We here determine whether any of the properties of FR~IIs determined in the
above analysis are correlated with the FR~II environment.  In
\S\ref{subsec:correlation} we examine correlations of jet power with
stellar luminosity, BCG velocity dispersion, cluster richness, and FR~II
lifetime.  The impact of our assumptions about the ICM model used in
\S\ref{subsec:correlation} is assessed in \S\ref{subsec:icm_variations}.
We finally examine the broader relationship between the FR~II and radio
galaxy fractions and galaxy cluster richness in \S\ref{subsec:friifrac}.

\subsection{Correlations with Jet Power}
\label{subsec:correlation}

Radio jets are powered by the accretion of gas onto nuclear, supermassive
black holes (SMBHs).  Although the mechanism by which infalling gas is
expelled and collimated from the SMBH is not well understood, most jet
models rely on some variant of the Blandford-Znajak (BZ) mechanism
\citep{blandford77}.  The BZ mechanism explains jet outflows from SMBHs by
postulating a magnetic field anchored to an accretion disc around a rapidly
spinning SMBH.  As the accretion disk rotates around the SMBH, the field
lines become wound around the spin axis.  Some charged particles falling
into the SMBH will then follow the field lines and escape along the spin
axis of the SMBH.  Jet models which apply the BZ mechanism find that the
power that can be extracted from the SMBH and injected into the jet is
roughly proportional to the accretion rate and the square of the SMBH spin
\citep{mcnamara11}.  Larger black holes should therefore generally produce
more powerful jets.  It is interesting to see whether we can recover this
relationship in our FR~II sample.  While direct measurements of SMBH masses
are not available at the distances in our sample, we can check for the
existence of a correlation between jet power and two proxies for SMBH mass:
BCG luminosity and BCG velocity dispersion.  We also search for a
correlation between jet power and cluster richness, which may be related to
the fuel source and thus the accretion rate, and between lifetime and
cluster properties.

Because the MaxBCG catalog is not complete and our FR~II sample is
flux-limited, selection effects could induce an artificial correlation
between jet power and BCG or cluster properties.  For instance, if the
MaxBCG catalog suffered from Malmquist bias, a distant FR~II in our sample
would have to be both luminous at radio wavelengths (and hence have a large
jet power) and be hosted by a luminous galaxy.  This would lead to a
concentration of FR~IIs with high jet power and high BCG optical luminosity
in the sample and thereby induce a correlation between the two properties.
Completeness tests based on mock catalogs indicate that the MaxBCG catalog
is principally biased against low-mass clusters since the member galaxies
of such clusters tend to be less significant overdensities, although there
is a slight bias against extremely high-mass clusters as well \citep[see
Figure 7 of][for the detailed completeness function of the MaxBCG
catalog]{koester07b}.  Low-mass clusters typically host lower-luminosity
BCGs, so we should expect some correlation between jet power and BCG
luminosity due to these selection effects.  To avoid this bias, we use the
volume-limited sample described in \S\ref{subsec:sample_measurements} to
search for correlations between measured properties of the FR~IIs and
properties of their hosts.  We note that while we do not remove the
selection biases of the MaxBCG catalog, selection biases must be present in
both variables to induce a correlation; mitigating the selection biases in
just one variable is sufficient to remove this correlation.

\subsubsection{The Correlation between Jet Power and Stellar Luminosity}
\label{subsubsec:Q_lum_correlation}

\begin{figure}
\centering
\includegraphics[width=8cm]{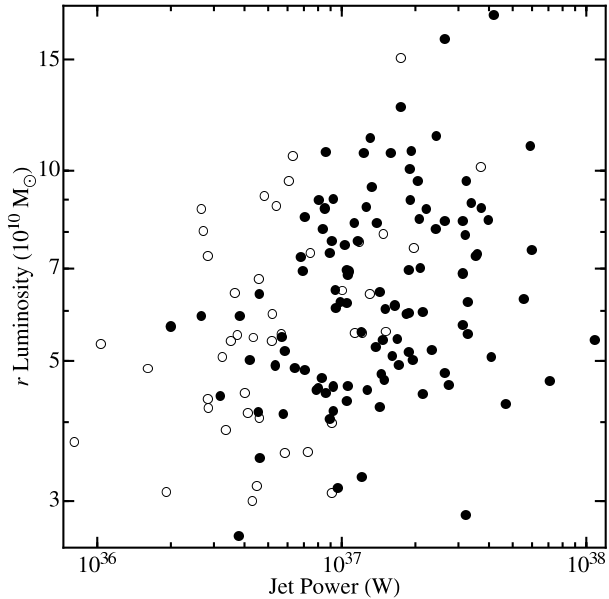}

\caption{The correlation between the $r$-band luminosity of the host BCG
and the estimated jet power.  Filled circles represent jets in the
volume-limited sample (see \S\ref{subsec:sample_measurements}) and open
circles represent jets in the full sample, but not the volume-limited
sample.  The correlation between the $r$-band luminosity and the jet power
for jets in the volume-limited sample is 0.284; there is a 0.3\% chance
that this correlation could be drawn from an uncorrelated population.  When
the jets from the full sample are included, the correlation becomes
stronger and more statistically significant.} \label{fig:rlum_q}

\end{figure}

The MaxBCG catalog provides a $K$-corrected $r$-band and $i$-band
luminosity for every BCG in the sample.  Because elliptical galaxies
generally exhibit very little star formation and have little dust, both the
$r$- and $i$-band luminosities should be well-correlated with the stellar
mass, which is correlated with the SMBH mass (\citealt{novak06} and
references therein).  If there is some power-law relationship between BCG
luminosity and SMBH mass, we expect a correlation between the logarithm of
the BCG luminosity and the logarithm of the jet power.  We calculate that
the Pearson correlation coefficient \citep{pearson1895} between $\log Q$
and $\log L_r$ is 0.284.  Given that our volume-limited sample has 106
degrees of freedom, the probability that a correlation coefficient of 0.284
or higher could be randomly drawn from uncorrelated data is only 0.3\% and
therefore statistically significant.  A correlation of similar strength
exists between the $i$-band luminosity and jet power.  Since jet power is
very strongly correlated to radio luminosity (see
\S\ref{subsec:q_measurement}), a correlation of similar statistical
significance exists between the radio luminosity of the radio lobes and the
optical luminosity of the host galaxy.  We show the relationship between
jet power and $r$-band luminosity in Figure \ref{fig:rlum_q}. 

\subsubsection{The Correlation between Jet Power and BCG Velocity
Dispersion} \label{subsubsec:Q_veldisp_correlation}

The velocity dispersions of elliptical galaxies are tightly correlated with
their nuclear SMBH masses \citep[e.g.,][]{gultekin09}.  If there is a
correlation between jet power and SMBH mass, we also expect a correlation
between jet power and host velocity dispersion.  We calculate the Pearson
correlation coefficient between $\log Q$ and $\log \sigma$ and find it to
be only 0.116.  There is a 33\% chance that this correlation could be drawn
from an uncorrelated population, and is therefore not statistically
significant.  This result is at odds with the correlation between jet power
and BCG stellar luminosity and may be due to two reasons.  First, SDSS does
not have spectra of the BCGs of all of the FR~IIs in our sample.  We only
have velocity dispersions for two-thirds of the BCGs in the volume-limited
sample, which reduces the statistical power of our sample.  Second, and
likely more important, is that although the stellar luminosities of
elliptical galaxies are strongly correlated with their velocity
dispersions, velocity dispersion is a weak function of luminosity
\citep[proportional to the fourth root of luminosity;][]{faber76}.  Our FR
II sample has a dynamic range in BCG luminosity of roughly an order of
magnitude.  Even if jet power and optical luminosity were perfectly
correlated, the resulting correlation between jet power and velocity
dispersion would exhibit a dynamic range of less than a factor of 2.  This
is small enough to be overwhelmed by moderate scatter.

\subsubsection{The Correlation between Jet Power and Cluster Richness}
\label{subsubsec:Q_richness}

Richer clusters tend to host larger BCGs that, in turn, tend to host larger
SMBHs.  Richer clusters also have more hot gas and many clusters have
cooling times shorter than 1 Gyr.  With larger SMBHs and a larger reservoir
of accretion material, it is plausible that cluster richness would be
correlated with jet power.  Alternatively, such a correlation could be
present due to a correlation between jet power and BCG luminosity paired
with a correlation between BCG luminosity and cluster richness.  We
calculate that the Pearson correlation coefficient between $\log Q$ and
$\log N_{ \textrm{gal}}$ is 0.094.  There is a 33\% probability that this
correlation could be drawn from an uncorrelated population and therefore it
is not statistically significant.  The non-detection is not too surprising
as we expect that this correlation would be weak at best since cluster
richness is a global property of the cluster and the jet power depends on
the microphysics governing the immediate vicinity of the SMBH.

\subsubsection{Correlations with FR~II Lifetime}
\label{subsubsec:lifetime_corr}

The FR~II lifetime is indicative of the length of time that a galaxy can
successfully supply its SMBH with cold gas.  Since the mechanism that
supplies this gas is not well understood, we use our data to determine if
FR~II lifetime is correlated with any BCG or cluster properties.  We test
for a correlation in two different ways.  The first method is identical to
the method for measuring correlations with BCG and cluster properties
explained above.  When we estimate the FR~II jet power (described in
\S\ref{subsec:q_measurement}), we simultaneously estimate a crude (due to
projection effects) FR~II age for every jet in our sample.  We then
calculate the Pearson correlation coefficient between these nominal ages
and various BCG and cluster properties (such as the BCG stellar
luminosity).  We find that none of the correlations between age and the BCG
and cluster properties discussed above are statistically significant.  The
major drawback of this method is that the age estimate is sensitive to the
length of the FR~II and, due to projection effects, the measured projected
length can be substantially smaller than the lobe length.  While we make
average corrections for projection effects, they nevertheless introduce
scatter into the relationship between the FR~II age and the various BCG and
cluster properties.  Furthermore, any individual FR~II is almost equally
likely to be detected at any point in its lifetime.  Since we use the FR
II's estimated age (not its lifetime), this will also introduce significant
scatter.

Our second approach is to split the sample into two bins at the median of
each parameter and test for correlations.  This provides two length
distributions for each parameter.  We perform the maximum likelihood
analysis described in \S\ref{subsec:maxlik} and fit these distributions to
the length distributions of the mock catalogs and see if there is a
statistically significant difference in lifetimes between FR~IIs in the two
bins.  However, we find no difference in lifetime between the two bins when
we split by stellar luminosity, BCG velocity dispersion, or cluster
richness.

\subsection{The Impact of the ICM}
\label{subsec:icm_variations}

The above analyses were all performed under the straightforward but
unrealistic assumption that the ICM is identical for all clusters in our
sample.  It is known, however, that richer clusters tend to harbor denser
ICMs with larger scale radii.  We therefore split our sample into two bins
at the median richness (14) to reexamine how strongly the FR~II jet power
and lifetime relate to the host cluster ICM.  The low richness bin has
richness values from 10 to 14.  We assign these clusters the intermediate
ICM model ($\chi_3$ in Table \ref{tbl:densities}) from \citet{jetha07},
which is derived from large groups and small clusters with a median
richness of nine.  Richer clusters (greater than 14) are assigned the
densest model ($\chi_5$ in Table \ref{tbl:densities}), which is the
geometric mean of the densities found by \citet{jones84} for clusters with
a median richness of 20.  

\begin{figure}
\centering
\includegraphics[width=8cm]{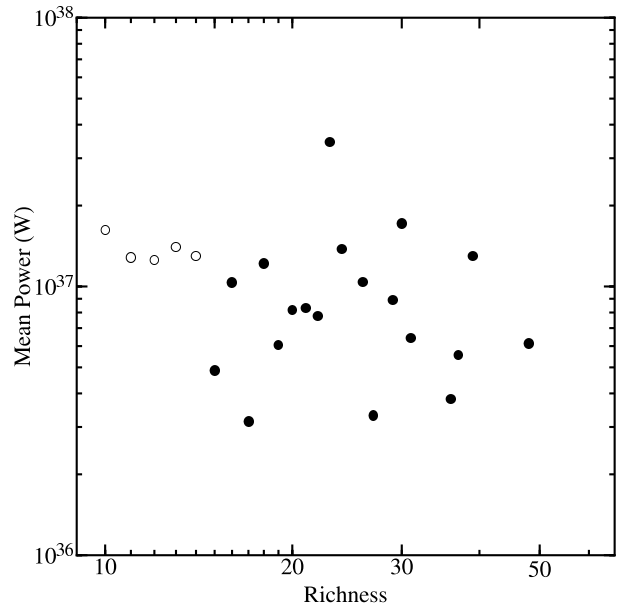}

\caption{The FR~II mean jet power as a function of host cluster richness
after applying a parameterization between cluster richness and ICM density.
Since larger clusters tend to have denser ICMs, we split our sample along
the median cluster richness of 14 and assume the $\chi_3$ ICM model (see
Table \ref{tbl:densities}) from \citet{jetha07} for the low-richness
clusters (open circles) and the $\chi_5$ ICM model from \citet{jones84} for
the high-richness clusters (filled circles).  We then estimate the jet
power of each FR~II with the inferred density as described in
\S\ref{subsec:q_measurement}.  The resulting correlation between jet power
and cluster richness, while statistically significant, is weak.}
\label{fig:density_param_q}

\end{figure}

For each FR~II we estimate the jet power with the method described in
\S\ref{subsec:q_measurement} and the density appropriate to the richness of
the cluster.  The power distribution as a function of richness is shown in
Figure \ref{fig:density_param_q}.  Although there is a correlation between
richness and FR~II jet power that is statistically significant at the
two-sigma level, it is weak.  If we simply compare the mean power of all
jets less than or equal to the median richness with those greater than the
median richness, we find that the jets in richer clusters are, on average,
less powerful by a factor of 2.  Although this density parameterization is
somewhat artificial because the ICM density is expected to steadily
continue to increase with richness, albeit with substantial scatter, and
spans an extreme range in density, this parameterization should be
sufficient to identify a strong correlation.  As the correlation between
inferred jet power and cluster richness is only marginal, both this result
and those in \S\ref{subsec:correlation} imply that the properties of FR IIs
are relatively insensitive to large-scale environmental factors.

\subsection{FR~II and Radio Galaxy Fraction vs.~Cluster Richness}
\label{subsec:friifrac}

If radio-mode feedback can prevent substantial gas cooling in clusters,
there may be an increase in duty cycle, lifetime, and jet power with
cluster richness.  In the previous subsections, we showed there is no
strong correlation between jet power or lifetime and richness.  Here we
investigate differences in duty cycle with the fraction of clusters with an
FR~II source and the fraction of clusters with a radio source of any kind.
We calculate the FR~II fraction in richness bins that have been corrected
for undetected sources.  This process would eliminate any potential
spurious correlations between FR~II fraction and cluster richness due to
the correlation between cluster richness and redshift.  

To calculate the fraction of all radio sources we cross-correlated the
FIRST and MaxBCG catalogs.  Rather than select radio sources at least
$10.\!\!^{\prime \prime}8$ (two FIRST resolution elements) from the BCG as
for FR~IIs, we selected BCGs with a radio source within $10.\!\!^{\prime
\prime}8$.  To mitigate Malmquist bias, we furthermore selected only those
radio sources whose radio luminosity was greater than $1.7 \times 10^{22}$
W Hz$^{-1}$ at 1.4 GHz.  This corresponds to the lowest luminosity that
FIRST can detect at $z = 0.3$, the largest redshift in the MaxBCG catalog.
\citet{bird08} performed a similar analysis using two additional, more
stringent luminosity cutoffs and found no substantive difference in the
relationship between radio fraction and cluster richness or luminosity.  We
therefore only use this luminosity cutoff.

\begin{figure}
\centering
\includegraphics[width=8cm]{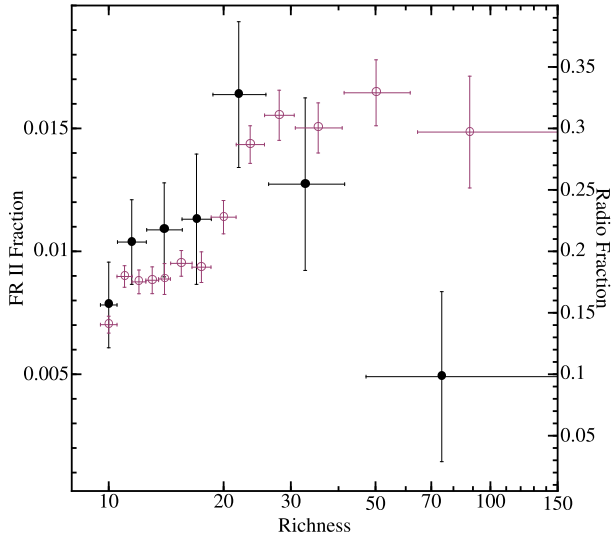}

\caption{The FR~II fraction (filled circles) and fraction of all radio
sources (open circles) as a function of cluster richness in the MaxBCG
catalog.  We define radio sources to be BCGs with radio emission detected
by FIRST within $10.\!\!^{\prime \prime}8$ (i.e.~two FIRST resolution
elements).  The horizontal error bars are the bin widths and the vertical
error bars are the binomial errors.} \label{fig:friifrac}

\end{figure}

\begin{figure}
\centering
\includegraphics[width=8cm]{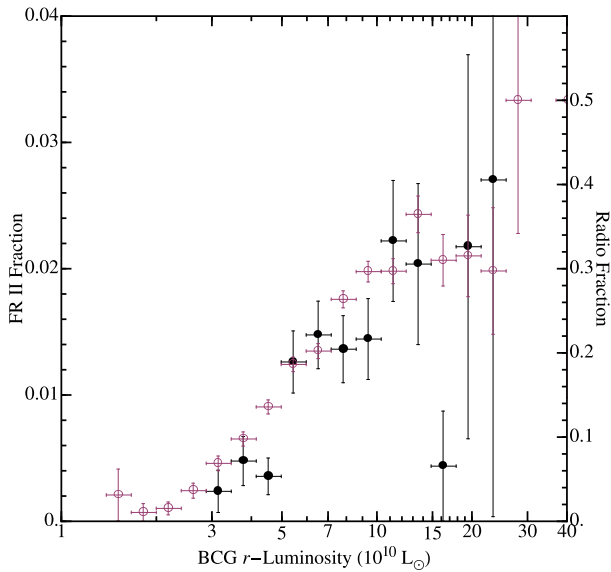}

\caption{The FR~II fraction (filled circles) and radio fraction (open
circles) as a function of the BCG $r$-band luminosity.  Radio sources are
defined as in Figure \ref{fig:friifrac}.  The horizontal error bars are the
bin widths and the vertical error bars are the binomial errors.}
\label{fig:friilumfrac}

\end{figure}

\begin{figure}
\centering
\includegraphics[width=8cm]{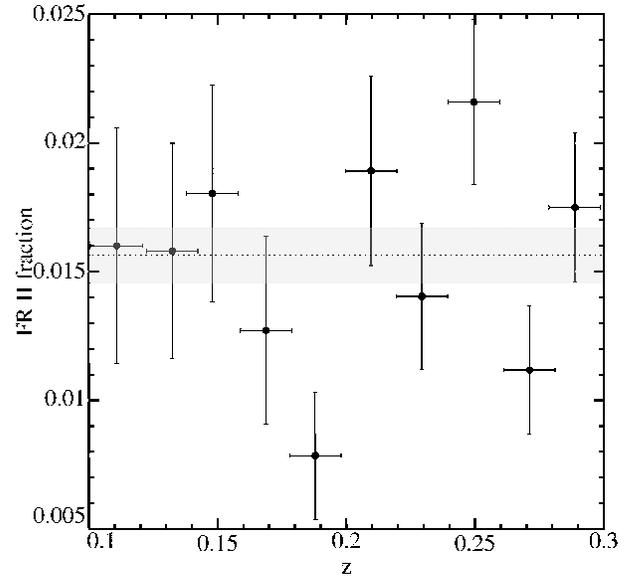}

\caption{The FR~II fraction of the volume-limited sample across the
redshift range of the MaxBCG catalog.  Horizontal error bars reflect bin
widths and vertical error bars reflect the binomial error on the FR~II
fraction in each bin.  The mean FR~II fraction across the entire sample is
shown by the dotted line with its binomial error represented by the shaded
region.  There is no evolution in the FR~II fraction over the redshift
range of the MaxBCG catalog.}\label{fig:z_frac}

\end{figure}

The FR~II and radio fractions as a function of cluster richness are shown
in Figure \ref{fig:friifrac}.  There is an increase in radio fraction by a
factor of 1.5 -- 2 with richness.  Although the statistics on the FR~II
fraction are weaker due to its smaller sample size, there appears to be a
corresponding trend in FR~II fraction with cluster richness.  We perform a
similar analysis to examine the relationship between the FR~II and radio
fractions with BCG luminosity and find that both fractions increase
strongly with BCG luminosity.  These relationships are shown in Figure
\ref{fig:friilumfrac}.  The increase in the radio fraction with stellar
luminosity is similar to the increase in the radio fraction with stellar
mass shown by \citet{best07} and \citet{stott12}.  We also find that the FR
II fraction does not evolve over the range of redshifts in the MaxBCG
catalog.  We show the FR~II fraction across the redshift range of the
volume-limited sample in Figure \ref{fig:z_frac}.


\section{Discussion}
\label{sec:discussion}

\subsection{Comparison to FR~II Lifetime Measurements in the Literature}
\label{subsec:comparison}

Our measurement of the FR~II lifetime is most directly comparable to the FR
II lifetime measurement by \citet{bird08} as their approach is largely
identical to our own.  They also modeled the observed FR~II length
distribution with mock catalogs to account for selection and projection
effects and used the KDA model to determine the length and luminosity
evolution.  \citet{bird08} found an FR~II lifetime of 1.5 $\times 10^7$ yr,
which is an order of magnitude below our value of 1.9 $\times 10^8$ yr.
There are three major differences between \citet{bird08} and this work that
together account for this difference.  The first is that the FR~IIs in the
\citet{bird08} sample were selected from a catalog of galaxy groups rather
than galaxy clusters.  Since the host group richness was much smaller,
\citet{bird08} took the $\chi_1$ ICM model as their default rather than the
$\chi_3$ model, which has a factor of two larger density parameter.  The
second major difference is that \citet{bird08} assumed jet power
distributions drawn from the literature rather than deriving the jet power
distribution from their (smaller) sample.  \citet{bird08} used jet power
distributions from \citet{blundell99} and \citet{sadler02} and found that
the lifetime is insensitive to these two choices.  This is because the
lifetime calculation depends more strongly on the median value of the jet
power distribution than on the higher-order moments; since the Blundell and
Sadler distributions have similar median powers, \citet{bird08} point out
that it is unsurprising that the two distributions produce similar
lifetimes despite their different shapes.  However, these distributions
were derived from samples of much more luminous FR~IIs than those in either
the \citet{bird08} sample or our own; FR~IIs in \citet{blundell99} were
selected from the 3C, 6C, and 7C catalogs and the \citet{sadler02}
distribution was derived from the bright end of the luminosity function.
Consequently, \citet{bird08} overestimated the median jet powers of the FR
IIs in their sample by approximately two orders of magnitude, and this led
to an underestimate of the lifetime by a factor of approximately five.  The
final difference between the two studies is our use of an updated
coefficient $c_1$ to calculate the lobe length (described in detail in
\S\ref{subsec:model_properties}).  The updated coefficient is smaller than
the original coefficient $c_1$ used by \citet{bird08} by about 40\% and
this increases the lifetime estimates by about 70\%.  When the differences
between the two studies in jet power distribution and $c_1$ are taken into
account, they almost entirely explain the order-of-magnitude discrepancy in
the lifetime measurement.  The remaining difference between the two
measurements is less than a factor of two and can be due to uncertainties
in the two measurements, or a small intrinsic difference in the FR~II
lifetime from group to cluster environments, or both.

\citet{blundell99} estimated the maximum lifetime of FR~IIs from their
observed distribution in the $P$-$D$ plane.  They noted that if the
lifetime were too short, one would predict far fewer long FR~IIs than are
observed and there would be a sharp maximum-size cutoff.  However, if the
maximum lifetime were too large, there would be no effect on the
distribution of points in the $P$-$D$ plane for a flux-limited sample since
FR~IIs above the true maximum lifetime would generally fall below the flux
limit; larger maximum lifetimes only decrease the predicted detection
fraction.  \citet{blundell99} found that the smallest maximum lifetime
consistent with their observations was 5 $\times$ 10$^8$ yr, in reasonable
agreement with our value.  The similarities in these lifetime estimates are
interesting because our study probes FR~IIs with jet powers approximately
two orders of magnitude below those in the sample of \citet{blundell99}.
\citet{wang08} fit a very similar sample with a typical maximum lifetime of
a few $\times$ 10$^7$ yr, although that choice was motivated by the
\citet{bird08} value. 

\citet{o'dea09} found FR~II lifetimes of a few $\times$ 10$^{6-7}$ yr with
the spectral index gradient method \citep{leahy89}.  They used a sample of
31 FR~IIs at the high end of the FR~II luminosity function (all of the FR
IIs in their sample are in the 3C catalog), whereas sources in our sample
are typically at much lower luminosities.  As they note, because their
flux-limited sample only includes the most powerful sources at high
redshift, these sources are expected to have the smallest lifetimes.  The
observed expansion velocities of the sources in the \citet{o'dea09} sample
are larger than the expansion velocities observed in our sample by about
two orders of magnitude.  This difference does not appear to be due to
differences in the two methods, however, since we obtain expansion
velocities consistent with those found by \citet{o'dea09} when we apply the
KDA model to the sources in their sample.  This suggests that the
discrepancy in the lifetime measurements between our work and
\citet{o'dea09} is due to the difference in jet powers between the two
samples rather than systematic discrepancies between the two methods.

Additionally, \citet{wang11} found that the length of FR~IIs is limited to
a maximum value by the entrainment of gas within the lobe.  \citet{wang11}
argued that beyond this maximum length the radio jet is disrupted,
transforming the FR~II into an FR~I.  Because more powerful jets reach this
maximum length more quickly, lifetime studies at the high end of the FR~II
luminosity function may systematically find lower lifetimes.

\subsection{FR~IIs as Sources of Heating in the ICM}
\label{subsec:icm_heating}

An outstanding problem in high-mass galaxy formation is the suppression of
substantial gas cooling.  As BCGs are almost always at the centers of
galaxy clusters, cool gas from the surrounding ICM is predicted to condense
onto the BCG.  Were this process to proceed uninterrupted, BCGs would grow
to be much larger than observed.  To quench these cooling flows, some
mechanism must exist to heat the ICM.

To first approximation, the heating of the ICM due to a radio lobe is
simply the enthalpy of the system,
\begin{equation}
\label{eq:enthalpy}
H = \frac{\gamma}{\gamma - 1}fpV,
\end{equation}
where $\gamma$ is the adiabatic index, $V$ is the volume of the lobe, $p$
is the pressure of the ICM immediately outside the lobe, and $f$ is an ad
hoc uncertainty factor \citep{birzan04, best07}.  Although we do not know
the pressure of the ICM outside the lobe, the KDA model does provide the
pressure inside the lobe (equation A4 of Kaiser \& Best, 2007).  Since it
is reasonable to assume that the lobe is overpressured relative to the ICM,
we calculate the enthalpy of the system with the lobe pressure and thereby
obtain an upper limit. 

To calculate the volume of the lobe, we assume a cylindrical geometry with
length $l$ and an axial ratio of $A = 4$.  The observed lengths of the FR
IIs are all extended by a factor of $\pi / 4$ to correct for projection
effects.  Finally, since the gas is relativistic, $\gamma = 4/3$.  The
enthalpy is therefore
\begin{equation}
\label{eq:observed_enthalpy}
H = \frac{256 p l^3}{\pi^2 A^2} f.
\end{equation}

We use the KDA model to calculate the lobe pressure for every FR~II in our
sample based on the estimated ages and powers of the jets (see
\S\ref{subsec:q_measurement}).  Although we may derive a lower limit on the
total ICM heating due to the adiabatic expansion of the lobe by dividing
the enthalpy by the estimated age of the FR~II, such an estimate is only
partially useful in determining whether FR~IIs can quench cooling flows in
galaxy clusters.  To be a successful mechanism to quench cooling flows, FR
IIs must not only provide a sufficient amount of total heat, but must also
provide a sufficient amount of heat throughout the central region of the
cluster.  Otherwise, cooling flows would be quenched in some regions and
would proceed unabated in others.  In particular, if gas on the equatorial
plane between the two lobes is not shocked, the assumption that the lobe
expands adiabaticly will provide an upper limit on the heating of gas in
this region.

\begin{figure*}
\centering
\includegraphics[width=16cm]{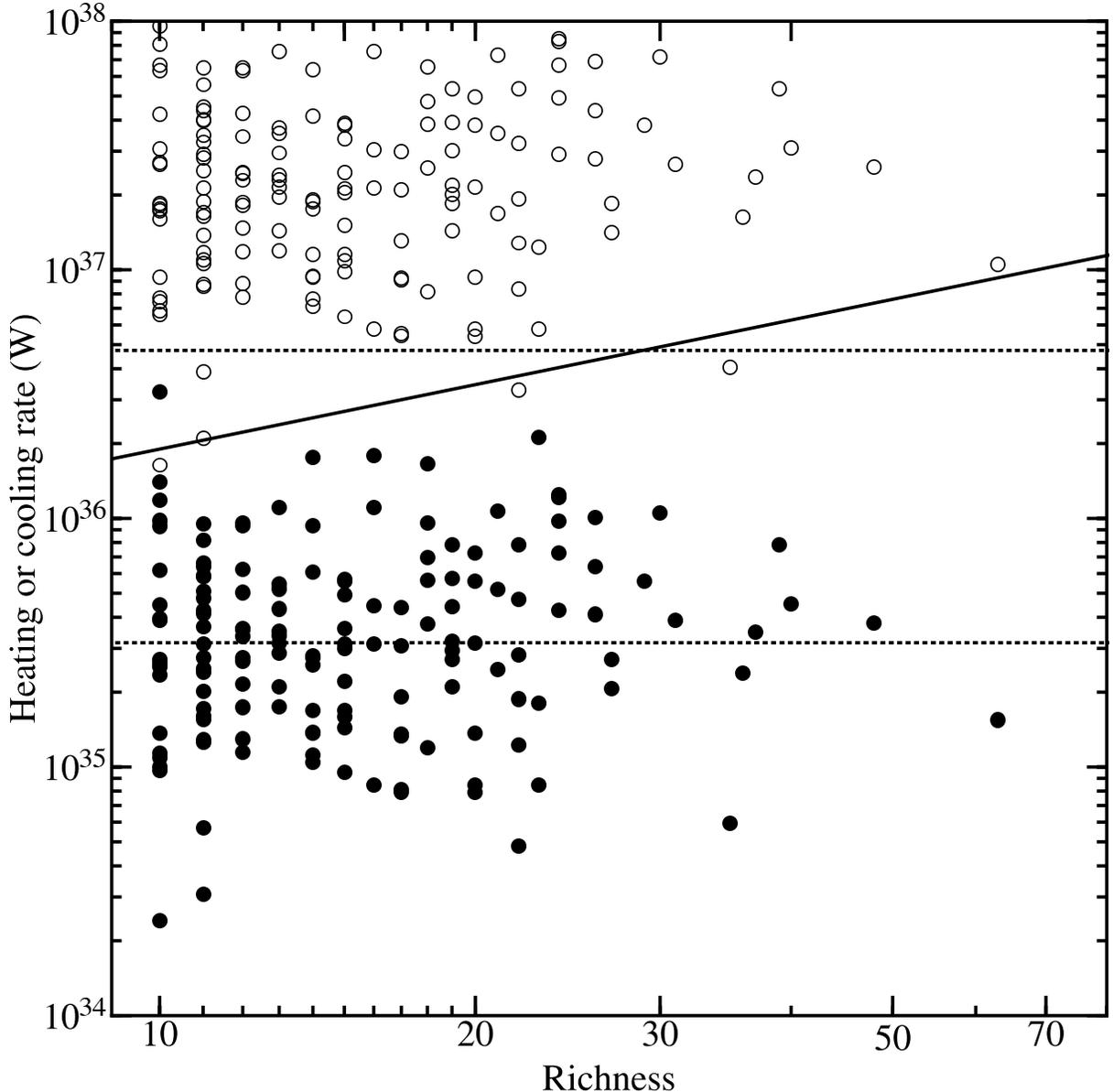}

\caption{The estimated ICM heating rates due to FR~IIs (filled circles) and
the estimated ICM cooling rates (solid line).  The heating rates represent
the enthalpy of each FR~II in our sample divided by the age of the FR~II.
Since the enthalpy is calculated from the pressure inside the lobe rather
than the external pressure, these points are upper limits on the average
heating rate.  The cooling rate (solid line) is calculated by combining the
scaling relation between X-ray luminosity and velocity dispersion from
\citet{oritz-gil04} with the scaling relation between velocity dispersion
and cluster richness from \citet{becker07}.  (A similar figure by
\citealt{best07} included several other scaling relations between X-ray
luminosity and velocity dispersion; since these are all very similar to the
Oritz-Gil relation, they are omitted here for clarity.)  The jet powers
(open circles) are higher than the required cooling rate, but only
$\sim$1\% of this power generates the enthalpy of the lobe.  Approximately
10\% of the jet power is needed to balance cooling.  (For comparison, the
open circles represent the jet powers of the FR~IIs in our sample,
estimated in \S\ref{subsec:q_measurement}).  The lower dotted line
represents the median heating of FR~IIs in our sample assuming an
uncertainty factor of $f = 1$ (i.e.~assuming that our estimate of the ICM
heating is exactly equal to the true ICM heating).  The upper dotted line
represents the median ICM heating assuming an uncertainty factor of $f =
15$.  This is approximately the value of $f$ required for FR~II heating to
balance the ICM cooling.}\label{fig:heating_cooling}

\end{figure*}

To determine whether ICM heating from the adiabatic expansion of the lobe
is sufficient to halt substantial gas cooling, we compare the average
heating rate to the average cooling of the ICM.  While we do not have X-ray
luminosities to directly estimate cooling rates, we can estimate the
typical cooling rate from the scaling relation between X-ray luminosity and
cluster velocity dispersion of
\begin{equation}
\label{eq:oritz-gil}
\log_{10}(L_{45}) = -1.34 + 2.0 \times \log_{10}(\sigma_{500}) \
\textrm{erg s}^{-1},
\end{equation}
where $L_{45}$ is the X-ray luminosity in units of $10^{45}$ erg s$^{-1}$
and $\sigma_{500}$ is the cluster velocity dispersion in units of 500 km
s$^{-1}$ \citep{oritz-gil04}.  We estimate the cluster velocity dispersions
from the scaling relation measured by \citet{becker07} of
\begin{equation}
\label{eq:becker}
\langle \ln \sigma \rangle = 6.17 + 0.436 \ln N_{\textrm{gal}}^{200} / 25. 
\end{equation}
These two equations provide an estimate of the ICM cooling in the MaxBCG
clusters.  The heating rate from the enthalpy of the lobes (assuming $f$ =
1) is shown with the estimated cooling rate in Figure
\ref{fig:heating_cooling}.  The typical heating rate due to the FR~II lobes
is at least an order of magnitude smaller than the estimated cooling rate,
implying that heating from radio-mode feedback from FR~IIs is insufficient
to quench cooling flows in rich clusters.  Figure \ref{fig:heating_cooling}
also demonstrates that the total jet powers exceed the amount required to
counteract cooling by approximately an order of magnitude.  That is, only
$\sim$10\% of the jet power is needed to counteract the predicted cooling,
while only $\sim$1\% of the jet power appears to contribute to the enthalpy
of the lobes.  These jets are consequently sufficiently powerful to
counteract cooling in individual clusters, but this power may not couple to
the ICM.  

There are several sources of uncertainty in the enthalpy calculation that
may bring the heating and cooling rates into better agreement.  First, our
calculation of the enthalpy of the lobe requires knowledge of the axial
ratio of the lobe.  We assume an axial ratio of $A = 4$, but because $H
\propto A^{-2}$, an overestimate of the axial ratio will lead to an
underestimate of the enthalpy.  However, the axial ratio can generally be
no smaller than $A = 2$, which corresponds to spherical lobes.  This would
increase the enthalpy by a factor of four.

Second, our calculation of the enthalpy of the lobe requires knowledge of
the pressure within the lobe.  This can be calculated from the KDA model,
but is dependent upon the density of the ICM outside the lobe.
Specifically, if the density profile of the ICM has a power-law slope of
$\beta = 2$, $p \propto \rho t^{-2}$, where $t$ is the lifetime of the FR
II \citep{kaiser07}.  Since $t \propto \rho^{1/3}$, we have that $p \propto
\rho^{1/3}$.  The enthalpy is proportional to $pV$, but the volume of the
lobe is an observable and is therefore independent of our assumptions about
the ICM density.  Thus,
\begin{equation}
\label{eq:enthalpy_scaling}
H \propto \rho^{1/3}.
\end{equation}
If the density profile does not have a power-law slope of $\beta = 2$,
there will be an additional dependence of the enthalpy on the jet power,
but if the slope is nearly 2, the deviation from the above relationship
will be small.  We therefore expect the dependence of the enthalpy on the
ICM density assumption to be weak.  If the typical ICM density is $5.9
\times 10^{-24}$ kg m$^{-3}$ at 260 kpc (the largest density we use in this
paper), rather than our assumed value of $7.2 \times 10^{-26}$ kg m$^{-3}$
at 391 kpc, the enthalpy increases by only a factor of 3.  

With a similar enthalpy analysis, \citet{best07} found that radio-mode
feedback has particular difficulty quenching cooling flows in very rich
clusters.  They estimated the average ICM heating due to the AGN with a
power-law relationship between radio luminosity and jet power from
\citet{best06}.  \citet{best07} compared this heating rate to the expected
cooling rate estimated from several generic scaling relations between X-ray
luminosity and velocity dispersion and found that the estimated heating
rate fell below the cooling rate derived from the X-ray luminosity for
clusters with a velocity dispersion above $\sigma_v \gtrsim 300$ km
s$^{-1}$.  All of the MaxBCG clusters are expected to be above this
velocity dispersion.

\citet{best07} identified three major sources of uncertainty in their
analysis: (1) the conversion from radio luminosity to jet power, (2) the
estimate of the ICM heating from the jet power, and (3) the fraction of ICM
heating that takes place within the cooling radius.  They parametrized
these uncertainties by the ad hoc uncertainty factor, $f$, in Equation
\ref{eq:observed_enthalpy}.  An uncertainty factor of $f = 1$ implies that
the ICM heating within the cooling radius is exactly the amount that
\citet{best07} infer from the radio luminosities.  Larger values of $f$
correspond to more ICM heating within the cooling radius than predicted,
and smaller values correspond to less heating.  Although \citet{best07}
argue that $f$ is likely close to unity, they note that if $f$ is somewhat
larger than 10 then the ICM heating due to radio-mode feedback would be
sufficient to quench cooling flows in nearly all galaxy clusters.  A better
constraint on $f$ could therefore determine whether radio-mode feedback is
a viable mechanism to quench cooling flows in rich clusters.  Because our
approach takes observed FR~II lengths and radio luminosities and uses an FR
II evolution model to constrain the jet power distribution of our sample,
we have effectively removed the first source of uncertainty from $f$.

As noted by \citet{best07}, there is additional uncertainty in the fraction
of the heating that takes place within the cooling radius.  We have assumed
that all of it does, but this fraction could be much less than 1.  If this
fraction is small, however, it becomes yet more difficult for radio-mode
feedback to quench cooling flows in rich clusters. 

Our results shown in Figure \ref{fig:heating_cooling} are similar to those
of \citet{best07}---radio-mode feedback can only quench cooling flows in
rich clusters ($N_{200} \geq 10$) if the uncertainty factor satisfies $f
\gtrsim 15$.  Meeting this criterion requires that the cores of galaxy
clusters be much denser than expected, that FR~II lobes be nearly
spherical, and that the ICM be only slightly underpressured relative to the
lobe.  This strongly constrains the environments in which radio-mode
feedback is an effective mechanism to quench cooling flows.  

Even if $f \gtrsim 15$, our duty cycle calculation suggests that FR~IIs can
be an effective ICM heating mechanism for only a small fraction of the
cluster's lifetime.  As described in detail in \S\ref{subsec:duty_cycle},
only $\sim$2\% of the BCGs in the MaxBCG catalog have FR~IIs after the
correction for selection effects.  If every BCG in the MaxBCG catalog is as
likely to host an FR~II as any other (a reasonable assumption given the
similarity between the vast majority of MaxBCG clusters in redshift and
richness), then FR~IIs are powered for $\sim$2\% of the cluster's lifetime.
If the uncertainty factor $f$ were greater than $\sim$15, FR~IIs would be
able to quench cooling flows during this $\sim$2\% of the cluster's
lifetime, but would be unable to quench cooling flows at any other time.
For FR~IIs to heat the ICM sufficiently to quench cooling flows for a
substantial fraction of the cluster's lifetime would require the
uncertainty factor to be yet an order of magnitude larger.  Since we have
largely eliminated the source of uncertainty (1) of the conversion from
radio luminosity to jet power, and the source of uncertainty (3) of the
fraction of ICM heating that takes place within the cooling radius only
acts to make $f$ smaller, the only way to increase $f$ is in the source of
uncertainty (2) of the estimate of the ICM heating from the jet power.  One
way to increase $f$ is if there were heating due to shocks near the cluster
center, as our calculations have only included ICM heating due to adiabatic
expansion.

Another possible solution that addresses both the apparently small value of
$f$ and the low duty cycle is if the FR~II continues to affect the ICM
after the radio source is no longer apparent.  Hydrodynamical simulations
by \citet{basson03} indicate that FR~IIs can substantially effect the
cooling properties of the ICM for significant period of time (up to an
order of magnitude longer than the lifetime of the jet) after the jet has
turned off by removing gas that otherwise would have cooled on longer
timescales.  Given a longer period of time in which to couple to the
external gas, FR~IIs could be more efficient than their lifetime alone
suggests.  Our results therefore imply that if FR~IIs are major sources of
heating in galaxy clusters, they contribute the vast majority of this
heating after their demise.

The range of jet powers of these FR~IIs provides some potential constraints
on the jet powers in the additional 20\% of the BCGs in the MaxBCG catalog
that host other cluster radio sources.  The two most likely scenarios for
why these other sources do not exhibit an FR~II morphology are that they
fall below the critical jet power criterion \citep[e.g.,][]{kaiser07} and
that local variations in the ICM have disrupted their jets.  With the
notable exception of Cygnus A \citep{carilli94}, nearly all of the early
evidence for the impact of radio sources on the ICM came from FR~Is such as
Perseus A \citep{boehringer93, fabian00} and Hydra A \citep{mcnamara00}.
If the vast majority of cluster radio sources are not FR~IIs because they
are insufficiently powerful, then their jet powers must fall below the FR
IIs in our sample, or $\lesssim 5 \times 10^{36}$ W.  This is at the low
end of the estimates of the power required to inflate cavities seen in
X-ray gas of $\sim 10^{35-39}$ W \citep[e.g.,][]{birzan08}, although the
clusters studied with X-rays may not be representative of all cluster radio
galaxies.

If the jet powers for these other radio sources are below the typical FR~II
jet powers by only one order of magnitude, or they are jets that exceed the
critical value but have been disrupted, then they will be sufficiently
powerful to counteract cooling in their clusters.  The first scenario would
require the coupling between jet power and the ICM to be closer to 100\%.
The second scenario could be constrained with detailed comparisons between
the cavity powers and lobe enthalpies of FR~Is and IIs.  In addition, while
our data suggest that the cavity power may be substantially less than the
jet power for FR~IIs, a different relation between cavity and jet power may
hold for FR~Is.  If they are more equal, then many FR~Is may be too weak to
create FR~IIs.  Better estimates of the environmental conditions and
intrinsic properties that produce the FR~I/II dichotomy would also provide
valuable new constraints on the properties of the radio sources that likely
ultimately produce feedback on the ICM.

\section{Summary}
\label{sec:summary}

We estimate the typical lifetime, jet powers, and ages for a new sample of
151 FR~IIs identified in data from the FIRST radio survey for clusters in
the MaxBCG catalog of \citet{koester07b}.  We use radio luminosities,
projected length measurements, and the KDA FR~II model to show that the
typical FR~II lifetime is $1.9 \times 10^8$ yr.  We find that the
distribution of FR~II jet power in BCGs is best described by a log-normal
distribution rather than a power-law distribution as has been previously
assumed.  We furthermore find that the jet power estimated for FR~IIs in
our sample is lower than that of other samples \citep[e.g.,][]{kaiser07,
bird08}.  The major uncertainties in the jet power and lifetime
measurements are the density of the ICM and the axial ratio of the jets.
Reasonable ranges in ICM density and axial ratio affect the lifetime
estimate at the factor of two level.

We examine the relationship between the properties of FR~IIs and the
properties of BCGs and the clusters.  We find that while BCG luminosity is
correlated with FR~II jet power, other cluster properties such as BCG
velocity dispersion and cluster richness are not strongly correlated with
either the jet power or the FR~II lifetime.  These results suggest that the
intrinsic properties of FR~IIs are not highly dependent on their
larger-scale environment.  

We evaluate the heating of the ICM from FR~IIs in galaxy clusters assuming
adiabatic expansion of overpressured radio lobes.  The time-averaged
enthalpy of the lobes implies that ICM heating due to FR~IIs is smaller
than ICM cooling by over an order of magnitude.  Although the jet provides
the lobe with a sufficient amount of energy to quench cooling flows, FR~IIs
cannot be major contributors to ICM heating without a viable mechanism by
which this energy can be distributed throughout the ICM.  This stands in
stark contrast to the total jet power, which \emph{exceeds} the cooling
rate by approximately an order of magnitude.  If the jet power could couple
with the ICM with $\sim$10\% efficiency, this would suffice to counteract
cooling, albeit for only the $\sim$2\% of BCGs that host an FR~II.  The
larger fraction of other radio sources in BCGs may be lower-power jets that
are insufficiently overpressured to form an FR~II morphology.  These
sources would require nearly 100\% coupling efficiency to the ICM and jet
powers only an order of magnitude below our sample to balance cooling. 

\acknowledgements

We would like to thank Brian McNamara, Chris O'Dea, Yang Wang, and the
anonymous referee for helpful comments on the manuscript.  We would also
like to thank Scott Gaudi for providing useful suggestions and acknowledge
Benjamin Shappee for discussing the manuscript with JMA.

This research makes use of the FIRST and NVSS radio surveys.  The National
Radio Astronomy Observatory is a facility of the National Science
Foundation operated under cooperative agreement by Associated Universities,
Inc.  The Cosmology Calculator by \citet{wright06} was used in the
preparation of this paper.

Funding for the SDSS and SDSS-II has been provided by the Alfred P.
Sloan Foundation, the Participating Institutions, the National Science
Foundation, the U.S. Department of Energy, the National Aeronautics and
Space Administration, the Japanese Monbukagakusho, the Max Planck
Society, and the Higher Education Funding Council for England. The SDSS
Web Site is http://www.sdss.org/.

The SDSS is managed by the Astrophysical Research Consortium for
the Participating Institutions. The Participating Institutions are
the American Museum of Natural History, Astrophysical Institute
Potsdam, University of Basel, University of Cambridge, Case Western
Reserve University, University of Chicago, Drexel University,
Fermilab, the Institute for Advanced Study, the Japan Participation
Group, Johns Hopkins University, the Joint Institute for Nuclear
Astrophysics, the Kavli Institute for Particle Astrophysics and
Cosmology, the Korean Scientist Group, the Chinese Academy of
Sciences (LAMOST), Los Alamos National Laboratory, the
Max-Planck-Institute for Astronomy (MPIA), the Max-Planck-Institute
for Astrophysics (MPA), New Mexico State University, Ohio State
University, University of Pittsburgh, University of Portsmouth,
Princeton University, the United States Naval Observatory, and the
University of Washington.

\bibliographystyle{apj}
\bibliography{/home/leavitt/antognini/bin/refs}

\end{document}